\documentclass[10pt]{article}
\pdfoutput=1
\usepackage{jheppub}
\usepackage{amsmath,amssymb,amsthm,latexsym,bbm,calc,wasysym,mathtools}
\usepackage[english]{babel}
\usepackage{graphicx,color}
\usepackage[dvipsnames]{xcolor}
\usepackage{tikz-cd}
\usepackage{tikz}
\usetikzlibrary{shapes.geometric}
\usetikzlibrary{arrows,matrix,calc,scopes,decorations.markings}
\usetikzlibrary{arrows.meta}
\usepackage{xspace}
\usepackage{pifont,dsfont}
\usepackage{marvosym}
\usepackage{slashed}
\usepackage{subfigure}
\usepackage{sidecap}
\usepackage{stmaryrd}
\usepackage{empheq}

\usepackage[export]{adjustbox}

\def\be{\begin{equation}}
\def\ee{\end{equation}}
\newcommand\sss{\scriptstyle}
\newcommand{\q}{\quad}

\newcommand{\mD}{\mathcal{D}}
\newcommand{\mH}{\mathcal{H}}
\newcommand{\mT}{\mathcal{T}}
\newcommand{\mC}{\mathcal{C}}

\newcommand{\ra}{\rangle}
\newcommand{\nn	}{\nonumber}

\newcommand{\wC}{\widetilde{C}}
\newcommand{\wR}{\widetilde{R}}

\newcommand{{\rd}}{\rm{d}}
\newcommand{\triCyl}{\mathbb{I}_3}
\newcommand{\smo}{\,{\sss \otimes}\,}
\newcommand{\smoc}{\,{\sss \otimes}_{\mathfrak{c}}\,}

\newcommand{\wa}{\widetilde{a}}
\newcommand{\wb}{\widetilde{b}}
\newcommand{\walpha}{\widetilde{\alpha}}
\newcommand{\wbeta}{\widetilde{\beta}}
\newcommand{\wm}{\widetilde{m}}
\newcommand{\wn}{\widetilde{n}}
\newcommand{\wip}{\widetilde{p}}
\newcommand{\wq}{\widetilde{q}}
\newcommand{\wc}{\widetilde{c}}
\newcommand{\wid}{\widetilde{d}}

\hypersetup{
	colorlinks=true,       
	linkcolor=BlueViolet,          
	citecolor=BrickRed,        
	filecolor=BrickRed,      
	urlcolor=BrickRed           
}

\tikzset{->1-/.style={decoration={
			markings,
			mark=at position #1 with {\arrow{Triangle[fill=black, width=4pt, length=4pt]}}},postaction={decorate}}}
\tikzset{->2-/.style={decoration={
			markings,
			mark=at position #1 with {\arrow{Triangle[fill=white, width=4pt, length=4pt]}}},postaction={decorate}}}
\tikzset{->3-/.style={decoration={
			markings,
			mark=at position #1 with {\arrow{Triangle[fill=white, width=4pt, length=4pt].Triangle[fill=white, width=4pt, length=4pt]}}},postaction={decorate}}}
\tikzset{->4-/.style={decoration={
		markings,
		mark=at position #1 with {\arrow{Triangle[fill=black, width=4pt, length=4pt].Triangle[fill=black, width=4pt, length=4pt]}}},postaction={decorate}}}
\tikzset{->5-/.style={decoration={
			markings,
			mark=at position #1 with {\arrow{Triangle[fill=lightgray, width=4pt, length=4pt]}}},postaction={decorate}}}			
\tikzset{->6-/.style={decoration={
			markings,
			mark=at position #1 with {\arrow{Triangle[fill=lightgray, width=4pt, length=4pt].Triangle[fill=lightgray, width=4pt, length=4pt]}}},postaction={decorate}}}			
\tikzset{->7-/.style={decoration={
			markings,
			mark=at position #1 with {\arrow{Triangle[fill=white, width=4pt, length=4pt].Triangle[fill=white, width=4pt, length=4pt].Triangle[fill=white, width=4pt, length=4pt]}}},postaction={decorate}}}
\tikzset{->8-/.style={decoration={
			markings,
			mark=at position #1 with {\arrow{Triangle[fill=black, width=4pt, length=4pt].Triangle[fill=black, width=4pt, length=4pt].Triangle[fill=black, width=4pt, length=4pt]}}},postaction={decorate}}}
\tikzset{->9-/.style={decoration={
			markings,
			mark=at position #1 with {\arrow{Triangle[fill=lightgray, width=4pt, length=4pt].Triangle[fill=lightgray, width=4pt, length=4pt].Triangle[fill=lightgray, width=4pt, length=4pt]}}},postaction={decorate}}}
				
\newcommand{\edge}[1]{
	\begin{tikzpicture}[scale=1,baseline=-0.2em]
	\coordinate (a) at (0,0);
	\coordinate (b) at (1,0);
	
	\draw[->1-=.6] (a) to (b);
	
	\node[above] at ($ (a) ! 0.5 ! (b) $) {{$\scriptstyle{#1}$}};
	\end{tikzpicture}
}

\newcommand{\edgeLong}[1]{
	\begin{tikzpicture}[scale=1,baseline=-0.2em]
	\coordinate (a) at (0,0);
	\coordinate (b) at (2,0);
	
	\draw[->4-=.58] (b) to (a);
	
	\node[above] at ($ (a) ! 0.5 ! (b) $) {{$\scriptstyle{#1}$}};
	\end{tikzpicture}
}

\newcommand{\edgeSub}[2]{
	\begin{tikzpicture}[scale=1,baseline=-0.2em]
	\coordinate (a) at (0,0);
	\coordinate (b) at (1,0);
	\coordinate (c) at (2,0);
	
	\draw[->1-=.6] (b) to (a);
	\draw[->2-=.6] (c) to (b);
	
	\node[draw,circle,scale=0.3, fill] at (b) {};
	
	\node[above] at ($ (a) ! 0.5 ! (b) $) {{$\scriptstyle{#1}$}};
	\node[above] at ($ (b) ! 0.5 ! (c) $) {{$\scriptstyle{#2}$}};
	\end{tikzpicture}
}

\newcommand{\edgeL}[1]{
	\begin{tikzpicture}[scale=1,baseline=-0.2em]
	\coordinate (a) at (0,0);
	\coordinate (b) at (1,0);
	
	\draw[->1-=.6] (b) to (a);
	
	\node[above] at ($ (a) ! 0.5 ! (b) $) {{$\scriptstyle{#1}$}};
	\end{tikzpicture}
}

\newcommand{\edgedef}[1]{
	\begin{tikzpicture}[scale=1,baseline=-0.2em]
	\coordinate (a) at (0,0);
	\coordinate (b) at (0.5,0);
	\coordinate (c) at (1,0);
	
	\draw[] (a) to [bend right=25] (b);
	\draw[->1-=.15] (b) to [bend left=25] (c);
	
	\node[above] at ($ (a) ! 0.5 ! (b) $) {{$\scriptstyle{#1}$}};
	\end{tikzpicture}
}

\newcommand{\tritri}[3]{
	\begin{tikzpicture}[scale=1,baseline=0.4em]
	\coordinate (a) at(0,0);
	\coordinate (b) at (1,0);
	\coordinate (c) at (0.5,0.866);
	
	\draw[->4-=.65] (b) to (a);
	\draw[->3-=.65] (c) to (b);
	\draw[->6-=.65] (c) to (a);

	\node[below] at ($ (a) ! 0.5 ! (b) $) {{$\scriptstyle{#1}$}};
	\node[right] at ($ (b) ! 0.5 ! (c) $) {{$\scriptstyle{#2}$}};
	\node[left] at ($ (c) ! 0.5 ! (a) $) {{$\scriptstyle{#3}$}};
	\end{tikzpicture}
}

\newcommand{\tritriSub}[5]{
	\begin{tikzpicture}[scale=1,baseline=0.4em]
	\coordinate (a) at(0,0);
	\coordinate (b) at (0.5,0);
	\coordinate (c) at (1,0);
	\coordinate (d) at (0.5,0.866);
	
	\draw[->1-=.65] (b) to (a);
	\draw[->2-=.65] (c) to (b);
	\draw[->3-=.65] (d) to (c);
	\draw[->6-=.65] (a) to (d);
	\draw[->5-=.5] (b) to (d);

	\node[below] at ($ (a) ! 0.5 ! (b) $) {{$\scriptstyle{#1}$}};
	\node[below] at ($ (b) ! 0.5 ! (c) $) {{$\scriptstyle{#2}$}};
	\node[right] at ($ (c) ! 0.5 ! (d) $) {{$\scriptstyle{#3}$}};
	\node[left] at ($ (d) ! 0.5 ! (a) $) {{$\scriptstyle{#4}$}};
	\node[right, node distance=-1em] at ($ (b) ! 0.2 ! (d) $) {{$\scriptstyle{#5}$}};			
	\end{tikzpicture}
}

\newcommand{\cyli}[2]{
	\begin{tikzpicture}[scale=1,baseline=0.4em]
	\coordinate (a) at(0,0);
	\coordinate (b) at (0.7,0);
	\coordinate (c) at (0.7,0.7);
	\coordinate (d) at (0,0.7);
	
	\draw[->1-=.65] (b) to (a) ;
	\draw[->3-=.75] (b) to (c);
	\draw[->1-=.65] (c) to (d);
	\draw[->2-=.65] (a) to (d);		
	
	\node[below] at ($ (a) ! 0.5 ! (b) $) {{$\scriptstyle{#1}$}};
	\node[left] at ($ (a) ! 0.5 ! (d) $) {{$\scriptstyle{#2}$}};
	\node[right] at ($ (b) ! 0.5 ! (c) $) {{ }};
	\node[above] at ($ (d) ! 0.5 ! (c) $) {{ }};
	\end{tikzpicture}
}

\newcommand{\doublecyli}[3]{
	\begin{tikzpicture}[scale=1,baseline=0.4em]
	\coordinate (a) at(0,0);
	\coordinate (b) at (0.7,0);
	\coordinate (c) at (0.7,0.7);
	\coordinate (d) at (0,0.7);
	\coordinate (e) at (1.4,0);
	\coordinate (f) at (1.4,0.7);

	\draw[->1-=.65] (b) to (a);
	\draw[->3-=.75] (b) to (c) ;
	\draw[->1-=.65] (c) to (d) ;
	\draw[->2-=.65] (a) to (d) ;
	\draw[->5-=.65] (e) to (b) ;
	\draw[->6-=.75] (e) to (f) ;
	\draw[->5-=.65] (f) to (c) ;

	\node[below] at ($ (a) ! 0.5 ! (b) $) {{$\scriptstyle{#1}$}};
	\node[left] at ($ (a) ! 0.5 ! (d) $) {{$\scriptstyle{#2}$}};
	\node[below] at ($ (b) ! 0.5 ! (e) $) {{$\scriptstyle{#3}$}};
	\node[right] at ($ (e) ! 0.5 ! (f) $) {{ }};
	\node[above] at ($ (c) ! 0.5 ! (f) $) {{ }};		
	
	\end{tikzpicture}
}

\newcommand{\twotor}[2]{
	\begin{tikzpicture}[scale=1,baseline=0.4em]
	\coordinate (a) at(0,0);
	\coordinate (b) at (0.7,0);
	\coordinate (c) at (0.7,0.7);
	\coordinate (d) at (0,0.7);
	
	\draw[->1-=.65] (b) to (a) ;
	\draw[->2-=.65] (b) to (c);
	\draw[->1-=.65] (c) to (d);
	\draw[->2-=.65] (a) to (d);		
	
	\node[below] at ($ (a) ! 0.5 ! (b) $) {{$\scriptstyle{#1}$}};
	\node[left] at ($ (a) ! 0.5 ! (d) $) {{$\scriptstyle{#2}$}};
		\node[right] at ($ (b) ! 0.5 ! (c) $) {{ }};
	\node[above] at ($ (d) ! 0.5 ! (c) $) {{ }};
	\end{tikzpicture}
}

\newcommand{\thrice}[2]{
	\begin{tikzpicture}[scale=1,baseline=0.4em]
	\coordinate (a) at(0,0);
	\coordinate (b) at (0.7,0);
	\coordinate (c) at (0.35,0.606);

	\draw[->1-=.65] (b) to (a);
	\draw[->2-=.65] (c) to (b);
	\draw[->5-=.65] (a) to (c);
	
	\node[draw,circle,scale=0.3,fill] at (a) {};
	\node[draw,circle,scale=0.3,fill] at (b) {};
	\node[draw,circle,scale=0.3,fill] at (c) {};
	
	\node[right] at ($ (b) ! 0.5 ! (c) $) {{$\scriptstyle{#1}$}};
	\node[left] at ($ (c) ! 0.45 ! (a) $) {{$\scriptstyle{#2}$}};
	\end{tikzpicture}
}

\newcommand{\trivalent}[4]{
	\begin{tikzpicture}[scale=1,baseline=0em]
	\coordinate (a) at(0,0);
	\coordinate (b) at (0,0.7);
	\coordinate (c) at (-0.606,-0.35);
	\coordinate (d) at (0.606,-0.35);
	
	\draw[->1-=.65] (a) to (b);
	\draw[->2-=.65] (a) to (c);
	\draw[->5-=.65] (a) to (d);
	
	\node[above] at ($ (a) ! 0.3 ! (d) $) {{$\scriptstyle{#1}$}};
	\node[right] at ($ (a) ! 0.6 ! (b) $) {{$\scriptstyle{#2}$}};
	\node[above] at ($ (a) ! 0.65 ! (c) $) {{$\scriptstyle{#3}$}};
	\node[below] at ($ (a) ! 0.5 ! (d) $) {{$\scriptstyle{#4}$}};
	\end{tikzpicture}
}

\newcommand{\triface}[3]{
	\begin{tikzpicture}[scale=1,baseline=0.4em]
	\coordinate (a) at(0,0);
	\coordinate (b) at (0.7,0);
	\coordinate (c) at (0.35,0.606);
	
	\draw[->1-=.65] (a) to (b);
	\draw[->2-=.65] (c) to (b);
	\draw[->5-=.65] (a) to (c);
	
	\node[right] at ($ (b) ! 0.5 ! (c) $) {{$\scriptstyle{#1}$}};
	\node[left] at ($ (c) ! 0.45 ! (a) $) {{$\scriptstyle{#2}$}};
	\node[below] at ($ (a) ! 0.45 ! (b) $) {{$\scriptstyle{#3}$}};
	\end{tikzpicture}
}

\newcommand{\basisOne}[6]{
	\begin{tikzpicture}[scale=1,baseline=0em]
	\coordinate (a) at(0,0);
	\coordinate (b) at (0.7,0);
	\coordinate (c) at (0.35,0.606);
	\coordinate (d) at (0,-0.7);
	\coordinate (e) at (0.7,-0.7);
	\coordinate (f) at (-0.606,0.35);
	\coordinate (g) at (-0.256,0.956);
	\coordinate (h) at (1.306,0.35);
	\coordinate (i) at (0.956,0.956);
	
	\draw[->1-=.65] (b) to (a);
	\draw[->2-=.65] (c) to (b);
	\draw[->5-=.65] (a) to (c);
	\draw[->4-=0.75] (a) to (d);
	\draw[->8-=0.85] (e) to (d);
	\draw[->4-=0.75] (b) to (e);
	\draw[->6-=0.75] (a) to (f);
	\draw[->9-=0.85] (f) to (g);
	\draw[->6-=0.75] (c) to (g);
	\draw[->7-=0.85] (i) to (h);
	\draw[->3-=0.75] (b) to (h);
	\draw[->3-=0.75] (c) to (i);

	\node[draw,circle,scale=0.3,fill] at (a) {};
	\node[draw,circle,scale=0.3,fill] at (b) {};
	\node[draw,circle,scale=0.3,fill] at (c) {};
	
	\node[below] at ($ (d) ! 0.5 ! (e) $) {{$\scriptstyle{#2}$}};
	\node[left] at ($ (a) ! 0.5 ! (d) $) {{$\scriptstyle{#1}$}};
	\node[right] at ($ (h) ! 0.6 ! (i) $) {{$\scriptstyle{#4}$}};
	\node[below] at ($ (b) ! 0.6 ! (h) $) {{$\scriptstyle{#3}$}};
	\node[left] at ($ (f) ! 0.7 ! (g) $) {{$\scriptstyle{#6}$}};
	\node[above] at ($ (g) ! 0.7 ! (c) $) {{$\scriptstyle{#5}$}};
	\end{tikzpicture}
}

\newcommand{\threeCyl}[3]{
	\begin{tikzpicture}[scale=1,baseline=1.2em]
	\coordinate (a) at(0,0);
	\coordinate (b) at (1,0);
	\coordinate (c) at (1,1);
	\coordinate (d) at (0,1);
	\coordinate (e) at (-0.40,0.40);
	\coordinate (e1) at (-0.10,0.40);
	\coordinate (e2) at (0.1,0.40);
	\coordinate (f) at (-0.40,1.40);
	\coordinate (g) at (0.6,1.4);
	\coordinate (g1) at (0.6,1.1);
	\coordinate (g2) at (0.6,0.9);
	\coordinate (h) at (0.6,0.4);
	
	\draw[->1-=.6] (b) to (a);
	\draw[->6-=.65] (b) to (c);
	\draw[->1-=.6] (c) to (d);
	\draw[->3-=.65] (a) to (d);
	\draw[->2-=.65] (a) to (e);
	\draw[->3-=.65] (e) to (f);
	\draw[->2-=.65] (d) to (f);
	\draw[->1-=.6] (g) to (f);
	\draw[->5-=.65] (c) to (g);
	\draw[->5-=.65] (b) to (h);
	\draw[->6-=.8] (h) to (g2);
	\draw[] (g1) to (g);
	\draw[->1-=.7] (h) to (e2);
	\draw[] (e1) to (e);

	\node[below] at ($ (a) ! 0.5 ! (b) $) {{$\scriptstyle{#1}$}};
	\node[left] at ($ (a) ! 0.4 ! (e) $) {{$\scriptstyle{#2}$}};
	\node[left] at ($ (e) ! 0.5 ! (f) $) {{$\scriptstyle{#3}$}};
	\node[right] at ($ (b) ! 0.5 ! (c) $) {{}};
	\node[above] at ($ (f) ! 0.5 ! (g) $) {{}};
	\end{tikzpicture}
}

\newcommand{\doubleThreeCyl}[4]{
	\begin{tikzpicture}[scale=1,baseline=1.2em]
	\coordinate (a) at(0,0);
	\coordinate (b) at (1,0);
	\coordinate (c) at (1,1);
	\coordinate (d) at (0,1);
	\coordinate (e) at (-0.40,0.40);
	\coordinate (e1) at (-0.10,0.40);
	\coordinate (e2) at (0.1,0.40);
	\coordinate (f) at (-0.40,1.40);
	\coordinate (g) at (0.6,1.4);
	\coordinate (g1) at (0.6,1.1);
	\coordinate (g2) at (0.6,0.9);
	\coordinate (h) at (0.6,0.4);
	
	\coordinate (bx) at (2,0);
	\coordinate (cx) at (2,1);
	\coordinate (e1x) at (0.90,0.40);
	\coordinate (e2x) at (1.1,0.40);
	\coordinate (gx) at (1.6,1.4);
	\coordinate (g1x) at (1.6,1.1);
	\coordinate (g2x) at (1.6,0.9);
	\coordinate (hx) at (1.6,0.4);

	\draw[->1-=.6] (b) to (a);
	\draw[->6-=.65] (b) to (c);
	\draw[->1-=.6] (c) to (d);
	\draw[->3-=.65] (a) to (d);
	\draw[->2-=.65] (a) to (e);
	\draw[->3-=.65] (e) to (f);
	\draw[->2-=.65] (d) to (f);
	\draw[->1-=.6] (g) to (f);
	\draw[->5-=.65] (c) to (g);
	\draw[->5-=.65] (b) to (h);
	\draw[->6-=.8] (h) to (g2);
	\draw[] (g1) to (g);
	\draw[->1-=.7] (h) to (e2);
	\draw[] (e1) to (e);
	
	\draw[->7-=.75] (bx) to (b);
	\draw[->7-=.75] (cx) to (c);
	\draw[->4-=.65] (bx) to (cx);
	\draw[->4-=.8] (hx) to (g2x);
	\draw[] (g1x) to (gx);
	\draw[->7-=.95] (hx) to (e2x);
	\draw[] (e1x) to (h);
	\draw[->8-=.9] (bx) to (hx);
	\draw[->8-=.9] (cx) to (gx);
	\draw[->7-=.75] (gx) to (g);
	
	\node[below] at ($ (a) ! 0.5 ! (b) $) {{$\scriptstyle{#1}$}};
	\node[below] at ($ (b) ! 0.5 ! (bx) $) {{$\scriptstyle{#4}$}};
	\node[left] at ($ (a) ! 0.4 ! (e) $) {{$\scriptstyle{#2}$}};
	\node[left] at ($ (e) ! 0.5 ! (f) $) {{$\scriptstyle{#3}$}};
	\node[right] at ($ (b) ! 0.5 ! (c) $) {{}};
	\node[above] at ($ (f) ! 0.5 ! (g) $) {{}};
	\end{tikzpicture}
}

\newcommand{\threeTor}[3]{
	\begin{tikzpicture}[scale=1,baseline=1.2em]
	\coordinate (a) at(0,0);
	\coordinate (b) at (1,0);
	\coordinate (c) at (1,1);
	\coordinate (d) at (0,1);
	\coordinate (e) at (-0.40,0.40);
	\coordinate (e1) at (-0.10,0.40);
	\coordinate (e2) at (0.1,0.40);
	\coordinate (f) at (-0.40,1.40);
	\coordinate (g) at (0.6,1.4);
	\coordinate (g1) at (0.6,1.1);
	\coordinate (g2) at (0.6,0.9);
	\coordinate (h) at (0.6,0.4);
	
	\draw[->1-=.6] (b) to (a);
	\draw[->5-=.65] (b) to (c);
	\draw[->1-=.6] (c) to (d);
	\draw[->5-=.65] (a) to (d);
	\draw[->2-=.65] (a) to (e);
	\draw[->5-=.65] (e) to (f);
	\draw[->2-=.65] (d) to (f);
	\draw[->1-=.6] (g) to (f);
	\draw[->2-=.65] (c) to (g);
	\draw[->2-=.65] (b) to (h);
	\draw[->5-=.8] (h) to (g2);
	\draw[] (g1) to (g);
	\draw[->1-=.7] (h) to (e2);
	\draw[] (e1) to (e);
	
	\node[below] at ($ (a) ! 0.5 ! (b) $) {{$\scriptstyle{#1}$}};
	\node[left] at ($ (a) ! 0.4 ! (e) $) {{$\scriptstyle{#2}$}};
	\node[left] at ($ (e) ! 0.5 ! (f) $) {{$\scriptstyle{#3}$}};
	\node[right] at ($ (b) ! 0.5 ! (c) $) {{}};
	\node[above] at ($ (f) ! 0.5 ! (g) $) {{}};
	\end{tikzpicture}
}

\newcommand{\thriceThree}[3]{
	\begin{tikzpicture}[scale=1,baseline=0em]
	\coordinate (a) at(0,0);
	\coordinate (b) at (0.3,-0.4);
	\coordinate (c) at (1,0);
	\coordinate (c1) at (0.4,0);
	\coordinate (c2) at (0.2,0);
	\coordinate (d) at (0,1);
	\coordinate (e) at (0.3,0.6);
	\coordinate (f) at (1,1);

	\draw[->1-=.65] (a) to (b);
	\draw[->2-=.6] (b) to (c);
	\draw[->5-=.8] (c) to (c1);
	\draw[] (c2) to (a);
	\draw[->4-=.6] (a) to (d);
	\draw[->4-=.6] (b) to (e);
	\draw[->4-=.6] (c) to (f);
	\draw[->1-=.65] (d) to (e);
	\draw[->2-=.6] (e) to (f);
	\draw[->5-=.6] (f) to (d);
	
	\node[draw,circle,scale=0.3,fill] at (a) {};
	\node[draw,circle,scale=0.3,fill] at (b) {};
	\node[draw,circle,scale=0.3,fill] at (c) {};
	\node[draw,circle,scale=0.3,fill] at (d) {};
	\node[draw,circle,scale=0.3,fill] at (e) {};
	\node[draw,circle,scale=0.3,fill] at (f) {};
	
	\node[left] at ($ (a) ! 0.5 ! (d) $) {{$\scriptstyle{#3}$}};
	\node[left] at ($ (a) ! 0.7 ! (b) $) {{$\scriptstyle{#1}$}};
	\node[below] at ($ (b) ! 0.75 ! (c) $) {{$\scriptstyle{#2}$}};

	\end{tikzpicture}
}

\title{\boldmath Excitation basis for (3+1)d topological phases}

\author[a,b]{Clement Delcamp,}

\affiliation[a]{Perimeter Institute for Theoretical Physics,\\ 31 Caroline Street North, Waterloo, Ontario  N2L 2Y5, Canada}
\affiliation[b]{Department of Physics $\&$ Astronomy and Guelph-Waterloo Physics Institute \\  University of Waterloo, Waterloo, Ontario N2L 3G1, Canada}

\emailAdd{cdelcamp@perimeterinstitute.ca}

\abstract{We consider an exactly solvable model in 3+1 dimensions, based on a finite group, which is a natural generalization of Kitaev's quantum double model. The corresponding lattice Hamiltonian yields excitations located at torus-boundaries. By cutting open the three-torus, we obtain a manifold bounded by two tori which supports states satisfying a higher-dimensional version of Ocneanu's tube algebra. This defines an algebraic structure extending the Drinfel'd double. Its irreducible representations, labeled by two fluxes and one charge,  characterize the torus-excitations. The tensor product of such representations is introduced in order to construct a basis for (3+1)d gauge models which relies upon the fusion of the defect excitations. This basis is defined on manifolds of the form $\Sigma \times \mathbb{S}_1$, with $\Sigma$ a two-dimensional Riemann surface. As such, our construction is closely related to dimensional reduction from (3+1)d to (2+1)d topological orders.  }

\begin{document} 
	\vspace*{-2em}
	\maketitle
	\flushbottom
	
	\newpage
\section{Introduction}

Over the past two decades, exotic quantum states such as \emph{gapped} systems with non-trivial topological order have attracted considerable attention, see {\it e.g.} \cite{wen1989vacuum, wen1989chiral, wen1990topological, Kitaev1997, Levin:2004mi, levin2012braiding, mesaros2013classification, Kong:2014qka}. Such topological phases display \emph{quasi-particle excitations} on top of a \emph{ground state} which is defined by a \emph{topological quantum field theory} (TQFT). In two dimensions, an interesting class of topological order is described by (2+1)d \emph{gauge} theories. In that case, the excitations are characterized by a quantum group constructed from the gauge group of the theory \cite{Bais:2002ny, Lan2013, 2015arXiv150201690K}. This means in particular that the quasi-particles are labeled by irreducible representations of the quantum group.

An example of (2+1)d gauge theory model of topological phases, referred to as the Twisted Quantum Double (TQD) model \cite{hu2013twisted, mesaros2013classification, Bullivant:2017qrv}, is provided by an Hamiltonian extension of the \emph{Dijkgraaf-Witten} topological theory \cite{dijkgraaf1990}. The excitations of this model are characterized by the \emph{twisted Drinfel'd double} $\mD^\alpha(G)$ of the finite group $G$, with $\alpha$ a 3-cocycle over $G$. In the case of a trivial $3$-cocycle, the TQFT reduces to $BF$ theory and the TQD model boils down to \emph{Kitaev's quantum double model} \cite{Kitaev1997}.

The lattice Hamiltonian of Kitaev's quantum double model yields \emph{magnetic} and \emph{electric} \emph{point-like} excitations, both supported by \emph{punctures}, where punctures are obtained by removing solid disks from the surface. In this context,  the \emph{twice-punctured two-sphere} (or cylinder) plays a special role for two reasons. Firstly, this is the simplest topology supporting excitations. Secondly, the gluing of two cylinders results in another cylinder, hence defining an algebra on the Hilbert space of states, referred to as \emph{Ocneanu's tube algebra} \cite{ocneanu1993, ocneanu2001}. By defining specific excited states on the cylinder, we can confirm explicitly that this algebra is equivalent to the Drinfel'd double $\mD(G)$ of the gauge group \cite{Lan2013, DGTQFT, DDR1, bultinck2017anyons}. The representation theory of $\mD(G)$ can then be used to define the so-called {\it fusion basis} \cite{KKR, Hu:2015dga, DGTQFT, DDR1}. This basis, which is labeled by representations (charges) and conjugacy classes (fluxes), can be defined for any punctured Riemann surface via a pant-decomposition.

As part of an ongoing attempt to understand in more detail (3+1)d topological phases \cite{Wang:2014oya, 2012FrPhy...7..150W, wang2014braiding, Moradi:2014cfa, Wan:2014woa, Bullivant:2016clk, Wen:2016cij, DD16, Dittrich:2017nmq, Williamson:2016evv, 2017arXiv170404221L, 2017arXiv170202148E, Riello:2017iti, Delcamp:2018wlb}, we propose in this paper a higher-dimensional extension of the fusion basis. The construction follows closely the (2+1)d one. The excitations are now supported by \emph{torus-boundaries} which arise from removing solid tori from a three-manifold. The equivalent of the cylinder is obtained by cutting open along one direction the three-torus. The resulting manifold, which is bounded by two tori, is the support of states which satisfy a $3$d generalization of Ocneanu's tube algebra. It turns out that the corresponding gluing operation yields an extension of the Drinfel'd double referred to as the {\it quantum triple} $\mT(G)$, whose definition is proposed in this paper. The representation theory of $\mT(G)$ can then be used, as in the (2+1)d case, to define a basis of excited states. While the (2+1)d fusion basis can be constructed for any punctured surface $\Sigma$, the extension we propose in this paper for (3+1)d models is defined on manifolds of the form $\Sigma \times \mathbb{S}_1$. In other words, we can think of this basis as a lifting of the (2+1)d fusion basis via a direct product with the circle $\mathbb{S}_1$. 

Therefore, our construction follows the same strategy as {\it dimensional reduction}, which is a technique widely used for the study of three-dimensional topological phases \cite{Moradi:2014cfa, wang2014braiding, Wang:2014oya, 2017arXiv170404221L, Tantivasadakarn:2017xbg}. This technique relies upon the compactification of one of the spatial directions into a small circle $\mathbb{S}_1$. The study of a (3+1)d topological order $\mathcal{C}^{\rm 3d}$ then boils down to studying several (2+1)d topological orders $\mathcal{C}^{\rm 2d}$. More precisely, in the case of a topological order $\mathcal{C}^{\rm 3d}_G$ described by a gauge theory with finite group $G$, we can symbolically write the dimensional reduction as $\mathcal{C}_G^{\rm 3d} = \bigoplus_C \mathcal{C}_{Z_C}^{\rm 2d}$, where $C$ is a conjugacy class of the full group $G$ and $Z_C$ the centralizer of a representative element of $C$. We shall see how the relation between the quantum triple and the Drinfel'd double is the algebraic translation of such dimensional reduction.  

\bigskip \noindent
The paper is organized as follows. In sec.~\ref{sec:model}, we review some basic facts about $BF$ theory in order to motivate the form of the lattice Hamiltonian. Such Hamiltonian is valid both in (2+1)d and (3+1)d. In sec.~\ref{sec:ocneanu}, we revisit Ocneanu's tube algebra and explain how the Drinfel'd double structure naturally emerges. After presenting the main properties of this algebraic structure, we construct the fusion basis for (2+1)d topological phases. Finally, in sec.~\ref{sec:three}, we extend the previous construction and reveal the algebraic structure of the (3+1)d  excitations, namely the quantum triple. The representation theory of the quantum triple is presented and then used to define a basis of excitations for gauge theory model of (3+1)d topological phases. The paper has also one appendix where the technical details are relegated. 

\vspace{2em}
\section{Lattice gauge theory model of topological phases \label{sec:model}}
In this section, we describe the lattice Hamiltonian of the model under consideration, starting with a brief summary of the corresponding continuum theory, namely $BF$ theory. This lattice Hamiltonian which describes in (2+1)d nothing else than Kitaev's quantum double model \cite{Kitaev:1997wr} is naturally extended to (3+1)d. 
\subsection{$BF$ theory}
We are interested in topological phases with \emph{defect} excitations whose ground state is described by a discretized $BF$ theory \cite{horowitz1989, Baez:1995ph} with finite groups. In order to motivate the form of the \emph{lattice Hamiltonian} which yields the excitations we are interested in, we shall briefly review the canonical analysis of the theory in the case where the gauge group is a Lie group $G$ whose Lie algebra is denoted by $\mathfrak{g}$. In $d$ dimensions, the action reads
\be
	S[e,\omega] = \int_{\mathcal{M}}\text{tr}(e \wedge F(\omega))
\ee 
where $\mathcal{M} = \Sigma \times \mathbb{R}$, $e$ denotes a $\mathfrak{g}$-valued $(d-2)$-form, $\omega$ a connection on a trivial $G$-bundle and $F={\rm d}\omega +\omega \wedge \omega$ its curvature. The $BF$ action displays two kinds of gauge symmetries. First, there is a local $G$-rotational symmetry
\begin{equation}
	\delta_\Lambda e = [e,\Lambda] \q , \q \delta_\Lambda \omega = {\rd}_{\omega} \Lambda
\end{equation}
with $\Lambda$ a $\mathfrak{g}$-valued $(d-3)$-form. Secondly, there is a translational symmetry parametrized by a $\mathfrak{g}$-valued 0-form $N$
\be
	\delta_N e = {\rd}_{\omega}N \q , \q \delta_N \omega = 0
\ee
which follows from the Bianchi identity ${\rd}_{\omega}F=0$. The phase space of this theory is parametrized by the pull-back of both the field $e$ and the connection $\omega$ to $\Sigma$, denoted by $A^i_a$ and $E^b_j$ in local coordinates, respectively. 
Canonical analysis of the action reveals the first class constraints:
\be
	D_b E^b_j = 0 \q , \q F^i_{ab}(A) = 0
\ee
which generate the local symmetries of the action. We shall refer to these two constraints as the Gau{\ss} constraint and the flatness constraint (or zero-flux condition), respectively.
The $\mathfrak{g}$-valued connection transforms under gauge transformation as
\be
	g \triangleright A_a = gA_ag^{-1}+ g \partial_a g^{-1} \; .
\ee 
Let $\ell$ be a piecewise analytic curve, the holonomy $h_\ell(A) \in G$ along $\ell$ in $\Sigma$ is given by the path ordered exponential
\be
	h_{\ell}(A) = \mathcal{P}{\rm exp}\Big(-\int_\gamma A \Big) \; .
\ee
The theory is discretized by defining a graph connection on a graph $\Gamma$ which is obtained by assigning a group element $h_\ell$ to every $\ell \subset \Gamma$. A gauge transformation acts on such holonomies according to
\be
	\label{Gauss_hol}
	g \triangleright h_\ell = g_{t(\ell)}h_\ell g_{s(\ell)}^{-1} \; ,
\ee
where $s(\ell)$ and $t(\ell)$ denote the source and target vertices of $\ell$, respectively. Furthermore, the flatness constraint states that for every closed holonomy $h$ running along a contractible cycle, one has $h=\mathbbm{1}_G$, with $\mathbbm{1}_G$ the group identity. In the following, we will work in this discrete setting and replace the Lie group by a finite group that we still denote by $G$.

\subsection{Moduli space and Hilbert space}

By definition, flat connections have \emph{non-trivial} holonomies only along \emph{non-contractible} cycles and therefore we can label the gauge field configurations by homeomorphisms of the fundamental group $\pi_1(\Sigma)$ to the finite group $G$. The configuration space is then given \cite{dijkgraaf1990} by the moduli space $\mathcal{V}$ of flat $G$-bundles over $\Sigma$
\be
	\mathcal{V} = {\rm Hom}(\pi_1(\Sigma),G)/G
\ee
where the group acts by conjugation. 

From now on and for the rest of this section, we will of focus on the (2+1)d case, however the construction will generalize straightforwardly to the (3+1)d case. Let $\Sigma_\mathsf{g}$ be a Riemann surface of genus $\mathsf{g}$ such that $\mathcal{M}= \Sigma_\mathsf{g} \times \mathbb{R}$.  A presentation of the \emph{fundamental group} is provided by the group elements $(g_i,h_i)_{i=1}^\mathsf{g}$ satisfying $\prod_{i=1}^\mathsf{g}[g_i,h_i]=\mathbbm{1}_G$. A flat $G$-bundle is then obtained by such a presentation up to conjugation. 
Note that taking the quotient by the action of $G$ is to enforce gauge invariance at a root node which acts as source and target node of all the cycles of $\pi_1(\Sigma_\mathsf{g})$. Upon quantization of $BF$ theory on the space-time $\Sigma_{\mathsf{g}}\times \mathbb{R}$, the Hilbert space $\mH_{\Sigma_{\mathsf{g}}}$ of gauge-invariant functionals on the space of flat connections on $\Sigma_\mathsf{g}$ is introduced. It is well-known \cite{dijkgraaf1990} that in the case of $BF$ theory, every point of the finite set $\mathcal{V}_\mathsf{g}$ gives rise to one independent quantum state and therefore 
\be
	{\rm dim} \, \mH_{\Sigma_\mathsf{g}} = |\mathcal{V}_\mathsf{g}| \; .
\ee 
This means in particular that the Hilbert space is spanned by states $|g_1,h_1,\cdots,g_\mathsf{g},h_\mathsf{g} \ra$, with the group elements defined up to simultaneous conjugation, such that $\prod_{i=1}^\mathsf{g}[g_i,h_i]=\mathbbm{1}$, which are in one-to-one correspondence with the flat $G$-bundles described above. Note that this equality is not true anymore in the case of the Dijkgraaf-Witten model which can be thought as a deformed version of $BF$ theory where the gauge invariance is twisted by a \emph{cohomology} class.  

More generally, let us consider a genus-$\mathsf{g}$ surface $\Sigma_\mathsf{g}^p$ which contains $p$ punctures. The surface $\Sigma^p_\mathsf{g}$ is a genus-$\mathsf{g}$ surface with one disk removed around each puncture. Additionally, we require the presence of one {\it marked point} located at the boundary of every such disks. Naturally, punctures introduce additional non-contractible cycles along which the corresponding holonomies can be non-trivial. Let $\Gamma$ be a minimal graph embedded in $\Sigma_\mathsf{g}^p$ which captures the loops of the fundamental group $\pi_1(\Sigma_\mathsf{g}^p)$ and such that for each puncture there is a vertex of $\Gamma$ coinciding with the marked point. The configuration space is completely characterized by the holonomies along the edges of $\Gamma$. The Hilbert space $\mH_\Gamma$ is then given by the set of gauge invariant functionals $\psi: G^E \rightarrow \mathbbm{C}$ with $E$ the number of edges on the minimal graph.

Furthermore, for every puncture, we decide to relax the Gau{\ss} constraint at every vertex coinciding with a marked point. Note that we could allow for more vertices at punctures (or more generally at the boundary) at which the Gau{\ss} constraint would also be relaxed. This would require introducing an equal number of additional marked points \cite{DDR1}. However, in this paper, there will always be a single marked point at each puncture (or more generally at each piece of boundary) and therefore only a single vertex per puncture at which the Gau{\ss} constraint is relaxed. 

The group action is therefore reduced to {\it bulk vertices} only and punctures are the support of both {\it magnetic} and {\it electric} point-like excitations. The purpose of the next section will be to compute the non-Abelian statistics of these topological excitations supported by punctures.

\subsection{Lattice Hamiltonian}
Let us now make our construction more explicit by introducing the corresponding lattice Hamiltonian. Recall that we are interested in the Hilbert space $\mH_{\Sigma_\mathsf{g}^p}$ of gauge invariant functionals on the space of flat connections, which can be represented by $\mH_\Gamma$, such that $\Gamma$ captures $\pi_1(\Sigma_\mathsf{g}^p)$. To each bulk vertex $v$ of $\Gamma$, we associate a projector $\mathbb{A}_v$ which realizes the projection onto gauge invariant states. To every face $f$ of $\Gamma$, we associate a projector $\mathbb{B}_f$ which enforces the zero-flux condition (or flatness constraint).  

Let us consider a three-valent vertex with all the edges outgoing. According to \eqref{Gauss_hol}, the action of $\mathbb{A}_v$ must read
\be
	\mathbb{A }_v \triangleright \trivalent{}{g_1}{g_2}{g_3} = \frac{1}{|G|}\sum_{h \in G}
	\trivalent{}{g_1h}{g_2h}{g_3h} \; .
\ee
More generally, it can be expressed as follows
\begin{align}
	\mathbb{A}_v = \frac{1}{|G|}\sum_{h\in G}
	\bigg( \bigotimes_{e: s(e)=v} R_h^e\bigg) \otimes
	\bigg(\bigotimes_{e: t(e)=v} L_h^e \bigg)
\end{align}
where $R_h$ and $L_h$ correspond to the right and the left group action, respectively, such that $R_h \triangleright \psi(g) = \psi(gh)$ and $L_h \triangleright \psi(g) = 
\psi(h^{-1}g)$. The operator $\mathbb{B}_f$ simply acts by multiplying the wave function with a delta function
\begin{align}
	\mathbb{B}_f \; \triangleright \psi(\{g\}) = \delta_{h_f,\mathbbm{1}_G}\, \psi (\{g\})  
\end{align}
where $h_f = \prod_{e \subset f}^{\leftarrow}g_e$ is the oriented product of the holonomies along the boundary of the face $f$. For instance, in the case of a triangular face, the action simply reads
\be
	\mathbb{B}_f \triangleright \triface{g}{h}{k} = \delta_{ghk^{-1},\mathbbm{1}_G}\triface{g}{h}{k} \; .	
\ee
The operators $\mathbb{A}_v$ and $\mathbb{B}_f$ commute \cite{Kitaev:1997wr, hu2013twisted} and the lattice Hamiltonian is finally given by

\begin{equation}
	\label{ham}
	\boxed{
	\mathbb{H} = - \sum_v^{} \mathbb{A}_v - \sum_f \mathbb{B}_f }
\end{equation}
which describes nothing else than Kitaev's quantum double model \cite{Kitaev:1997wr}. Let us now look at a specific example, namely the two-torus $\mathbb{T}_2$. By definition, we know that the Hilbert space $\mH_{\mathbb{T}_2}$ of gauge invariant functionals on the space of flat connections is spanned by states 
\be
	\label{Hibtor2}
	\mH_{\mathbb{T}_2} = 
	\big\{\frac{1}{|G|}\sum_{x \in G}|xgx^{-1},xhx^{-1} \ra_{\mathbb{T}_2} \; | \; gh=hg\big\} =: \Big\{ \twotor{g}{h} \Big\} 
\ee
where $|g,h \ra$ is a state defined on a graph capturing the two non-contractible cycles of $\mathbb{T}_2 = \mathbb{S}_1 \times \mathbb{S}_1$. We just introduced \eqref{Hibtor2} a graphical representation that we will now justify. Graph states can be chosen to be defined on the one-skeleton of a minimal discretization of the surface. The simplest discretization of the two-torus is provided by one parallelogram  whose opposite edges are identified. This discretization is made of one face on which $\mathbb{B}_f$ acts, two oriented edges, and one bulk vertex on which $\mathbb{A}_v$ acts. The two edges correspond to the non-contractible cycles. Furthermore, we decide to label with identical arrows (same shape and same color) edges which are identified. Making the identification of the edges explicit, we have the correspondence:
\begin{equation}
	\twotor{g}{h} \longleftrightarrow
	\includegraphics[scale=1, valign=c]{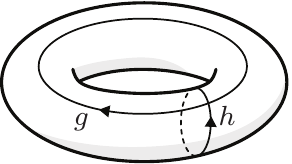} \; .
\end{equation}
In the following, every time we will make use of the graphical representation as in \eqref{Hibtor2}, it will be understood that the state is already projected so that gauge invariance is satisfied at every bulk vertex, and every magnetic flux going through a face associated with a contractible cycle is zero. In particular, when the labeling can be deduced from the identification of the edges or the zero-flux condition, it will often be left implicit. Let us examine carefully the case of the  basis states on the torus \eqref{Hibtor2}. First, the torus topology is encoded in the arrows decorating the edges. We can then deduce that there is a single bulk vertex, not located at a puncture, at which a group averaging is performed. Moreover, the zero-flux condition on the square face provides the delta function which enforces the commutation of the two holonomies. Thus, we have the equality
\begin{equation}
	\label{defRep2D}
	\twotor{g}{h} = \frac{\delta_{gh,hg}}{|G|}\sum_{x \in G}\twotor{xgx^{-1}}{xhx^{-1}}
\end{equation}
where both the delta function and the group averaging are redundant with the graphical representation, as such, it illustrates the projector property of both $\mathbb{A}_v$ and $\mathbb{B}_f$.

In the following, we will focus on states defined on the twice-punctured two-sphere $\mathbb{S}_2^2 \equiv \mathbb{I}$ which is obtained by cutting open the two-torus along one direction. Every flat bundle is trivial on the two-sphere. There is no non-contractible cycle. 
However, we introduce punctures which support both electric and magnetic excitations. Anywhere else, both constraints are satisfied. The twice-punctured two-sphere (which is topologically equivalent to a cylinder) $\mathbb{I}$ has a single non-contractible cycle. The discretization is now made of one face on which $\mathbb{B}_f$ acts, three edges, such that two of them define the boundary, and two vertices. Since the vertices, which are required to coincide with the marked points associated with the punctures, are now located at the boundary, we decide to relax the Gau{\ss} contraint which leads to electric excitations. As before, we have a graphical representation for the states spanning the Hilbert space $\mH_\mathbb{I}$:
\be
	\label{basisCyl2D}
	\mH_\mathbb{I} = 
	\Big\{ \cyli{g}{h} \Big\} \q {\rm with} \q
	\cyli{g}{h} \longleftrightarrow \includegraphics[scale=1, valign=c]{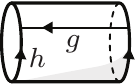} \; .
\ee
Recall that the states as represented above are already projected. Therefore, the zero-flux condition is enforced on the square face so that we can deduce the labeling for the edge decorated with a double white arrow, namely $g^{-1}hg$. However, it is clear from the graphical representation that, at the difference of the two-torus, there is no bulk vertex (not located at a puncture) on which $\mathbb{A}_v$ would act, and therefore there is no group averaging in the definition of the states \eqref{basisCyl2D}. Following the same strategy, we could define, for any surface $\Sigma_\mathsf{g}^p$, basis states in terms of holonomies labeling a graph capturing $\pi_1(\Sigma_\mathsf{g}^p)$. For instance, we will cover later the case of the thrice-punctured two-sphere.

\subsection{Equivalence relations}
Starting from a graph Hilbert spaces $\mH_\Gamma$ such that $\Gamma$ is embedded in $\Sigma_\mathsf{g}^p$, we obtain the Hilbert space $\mH_{\Sigma_\mathsf{g}^p}$ by identifying graph Hilbert spaces equivalent under deformation maps \cite{DGfluxQ, DGTQFT, DDR1}. More precisely, we identify states defined on different graphs if they are related by the following deformation maps: 
\begin{itemize}
	\item[$\circ$]{\it Changing orientation}---Two graph-states with opposite orientations and inverse group configurations are equivalent: 
	\be
	\edge{g} \sim \edgeL{g^{-1}}
	\ee
	\item[$\circ$]{\it Edge deformation}---Every edge can be freely deformed as long as the initial path and the resulting one are homotopy equivalent:
	\be
	\edge{g} \sim \edgedef{g}
	\ee
	\item[$\circ$]{\it Adding/removing vertices}---After subdivision of an edge, the Gau{\ss} constraint is enforced at the new vertex and the resulting graph-state equivalent to the original one is given by
	\be
	\edgeSub{g_1}{g_2} \sim \edgeLong{g_1g_2} \; ; \label{equivG}
	\ee
	or conversely, we can remove a bivalent vertex at which the Gau{\ss} constraint is imposed and the resulting group-labeling is the oriented product of the original ones. 
	\item[$\circ$]{\it Adding/removing edges}---After addition of an edge, a new closed face is created on which the operator $\mathbb{B}_f$ acts, hence enforcing the zero-flux condition. In the case of a triangular face, the resulting graph-state equivalent to the original one is given by
	\be
		\tritriSub{g_1}{g_2}{}{k}{} 
		\sim  \tritri{g_1g_2}{}{k} \; ; \label{equivF}
	\ee
	or conversely, we can remove an edge which is shared by closed faces on which the zero-flux condition is enforced.
\end{itemize}
In the following, we will consider the gluing of graph-states and use these equivalence relations in order to simplify the resulting states.

\vspace{2em}
\section{Ocneanu's tube algebra and Drinfel'd double \label{sec:ocneanu}} 
In the previous section, we introduced basis states for the Hilbert space $\mH_\mathbb{I}$ defined on the twice-punctured two-sphere. These cylinder states play a very important role in the characterization of elementary anyonic excitations. The fundamental reason is that the gluing of two cylinders gives another cylinder. Therefore, states defined on cylinders define an algebra called Ocneanu's tube algebra \cite{ocneanu1993, ocneanu2001}. In this section, we will define precisely the gluing procedure, show that this algebra actually corresponds to the Drinfel'd double $\mD(G)$ of the finite gauge group and present the main features of this rich algebraic structure. To do so, we will follow the steps of \cite{DDR1} where, to the best of our knowledge, the explicit definition of Ocneanu's tube algebra in the holonomy picture was first introduced. The representation theory of $\mD(G)$ will provide a natural way of constructing the so-called fusion basis for excited states. This derivation of known results will serve as a guideline for the (3+1)d generalization we propose in this paper.   
\subsection{Gluing of cylinders}

Starting from the simple observation that gluing two cylinders along a common boundary component leads to another cylinder, we will see that the gluing operation hides a well-known algebraic structure \cite{ocneanu1993, ocneanu2001, Lan2013, DDR1, bultinck2017anyons}. First, let us define more precisely this gluing operation \cite{Delcamp:2016eya}. This definition will also apply to the 3+1 case under consideration in the next section. Let $\mathcal{M}$ and $\mathcal{N}$ be two manifolds, $\partial \mathcal{M}$ and $\partial \mathcal{N}$ their boundary, which is a disjoint union of submanifolds, and $\mathcal{W}$ such a submanifold of both $\partial \mathcal{M}$ and $\partial \mathcal{N}$. We furthermore require $\mathcal{W}$ to be equipped with a marked point. The gluing of the manifold $\mathcal{M}$ and $\mathcal{N}$ along $\mathcal{W}$ is defined by identifying the boundary component $\mathcal{W}$ of both $ \partial \mathcal{M}$ and $\partial \mathcal{N}$ as well as the corresponding marked points. In the case of the cylinder, the boundary has two components, namely the two punctures, and the gluing of two cylinders consists in stacking them on top of each other. 

At the level of the graph-states, the gluing is performed by first identifying the edges and the vertices (associated to the marked points) located at the common boundary along which the gluing is performed. This identification procedure is denoted $\mathfrak{G}$. After identification, the vertices which once were located at boundaries are now bulk vertices at which the Gau{\ss} constraint must therefore be enforced using the operator $\mathbb{A}$. Similarly, in case the identification step $\mathfrak{G}$ produces new faces associated to contractible cycles, the zero-flux condition is enforced using the operator $\mathbb{B}$. Let $\Gamma_1$ and $\Gamma_2$ be two graphs embedded in the surfaces $\Sigma_1$ and $\Sigma_2$. The gluing operation of graph-states living in $\mH_{\Gamma_1}$ and $\mH_{\Gamma_2}$ is denoted $\star$ and is defined as
\be\label{starprod}
\begin{array}{cccccl}
	\star:& \;{\cal H}_{\Gamma_1} \otimes {\cal H}_{\Gamma_2}& \xrightarrow{{\;\; \mathfrak{G} \;\;}} & {\cal H}_{\text{aux}} & \xrightarrow{{\mathbb A} \,\circ\, {\mathbb B}} &{\cal H}_{{\Gamma_1 \cup \Gamma_2 / \sim}} \\
	& (\psi_1 , \psi_2) 
	&  \xmapsto{\phantom{\;\; \mathfrak{G} \;\;}} 
	& \mathfrak{G}(\psi_1 , \psi_2)& \xmapsto{\phantom{\mathbb{A} \, \circ \, \mathbb{B}}} 
	& \mathbb{A} \circ \mathbb{B} \triangleright \mathfrak{G} ( \psi_1,\psi_2)
\end{array}
\ee
where $\mH_{\rm aux}$ is the Hilbert space of functionals before enforcement of the constraints at the newly created bulk vertices and closed faces, and ${\Gamma_1 \cup \Gamma_2 / \sim}$ is the graph obtained after gluing of $\Gamma_1$ and $\Gamma_2$ up to equivalence relations. In the case of graph-states \eqref{basisCyl2D} defined on the cylinder, the computation goes as follows
\begin{align}
	\cyli{g_1}{h_1} \star \cyli{g_2}{h_2}
	 &= (\mathbb{A} \circ \mathbb{B}) \triangleright \mathfrak{G}
	 \Big( \cyli{g_1}{h_1}, \cyli{g_2}{h_2} \Big) \\
	 &=
	 \delta_{h_2,g_1^{-1}h_1g_1} \; 
	 \doublecyli{g_1}{h_1}{g_2} \sim 
	 \delta_{h_2,g_1^{-1}h_1g_1} \cyli{g_1g_2}{h_1} \; .
\end{align}
First, the two surfaces are identified which imposes a delta function between two holonomies, then the Gau{\ss} constraint is imposed at the four-valent vertex resulting from the gluing. The equivalence relations \eqref{equivG} and \eqref{equivF} can then be used to first remove the edge labeled by $h_2$ which leaves a two-valent vertex which can in turn be removed. We summarize this operation as follows, where it is understood that the result is up to equivalence relations: 
\begin{equation}
	\label{gluingCyl}
	\boxed{
	\cyli{g_1}{h_1} \star \cyli{g_2}{h_2} = \delta_{h_2,g_1^{-1}h_1g_1} \cyli{g_1g_2}{h_1}}
\end{equation}
which we recognize as the multiplication rule of the Drinfel'd double. 

\subsection{Drinfel'd double $\mD(G)$}

For a finite group $G$, the Drinfel'd double $\mathcal{D}(G)$ is an example of quasi-triangular Hopf algebra. We will not provide here a detailed description of this algebraic structure, only some of its main features. In particular, we will focus on the Hopf algebra structure and leave aside the quasi-triangularity property which describes the braiding of the corresponding anyonic excitations. A  detailed construction can be found in \cite{drinfel1988, Dijkgraaf1991}, see also \cite{DDR1} for many useful identities. 

As a Hopf algebra, the Drinfel'd double is a bialgebra obtained as a tensor product of an algebra and its dual coalgebra with opposite comultiplication, together with an antipode map $S$. A bialgebra $A$ over a field $k$ is a tuple $(A, \star, \mathbbm{1}, \Delta, \epsilon)$, such that $(A, \star, \mathbbm{1})$ is
an algebra over $k$ with multiplication $\star: A \otimes A \rightarrow A$ and unit $\mathbbm{1}: k \rightarrow A$, and $(A,\Delta,\epsilon)$ is a coalgebra over $k$ with
comultiplication $\Delta: A \rightarrow A \otimes A$ and a counit $\epsilon: A \rightarrow k$, such that $\Delta$ and $\epsilon$ are algebra homomorphisms. The antipode $S$ is an antihomomorphism such that 
\be
	\star \circ \, (\rm{id} \otimes S)\circ \Delta = \star \circ (S \otimes {\rm id}) \circ \Delta = \mathbbm{1} \circ \epsilon\; . 
\ee

Let us now make all these definitions explicit in the case of the Drinfel'd double. As a vector space, the Drinfel'd double is isomorphic to 
\be
	\mathcal{D}(G) \simeq \mathbb{C}[G] \otimes \mathcal{F}(G)
\ee
where $\mathbb{C}[G]$ is the group ring and $\mathcal{F}(G)$ is the Abelian algebra of linear functions on $G$. A basis for $\mD(G)$ is therefore provided by $\{g \smo \delta_h\}_{g,h \in G}$ where $\delta_h({\sss \bullet}) \equiv \delta(h,{\sss \bullet}) \equiv \delta_{h,{\sss \bullet}}$ is the Kronecker delta function supported on $h$. 

As a Hopf algebra, the Drinfel'd double comes equipped with the maps:
\begin{itemize}
	\item[$\circ$]{\it Multiplication:}
	\be
		\label{mulDrin}
		(g_1 \smo \delta_{h_1})\star (g_2 \smo \delta_{h_2}) := \delta_{h_1,g_1 h_2 g_1^{-1}}(g_1g_2 \smo \delta_{h_1})
	\ee
	with corresponding unit element $\mathbbm{1}_{\mD(G)}=\sum_{h \in G} \mathbbm{1}_G \smo \delta_h$. 
	\item[$\circ$]{\it Comultiplication:}
	\be
		\Delta(g \smo \delta_h) := \sum_{x,y \in G \atop xy = h}
		(g \smo \delta_x) \otimes (g \smo \delta_y) 
	\ee	
	with corresponding counit map $\epsilon(g \smo \delta_h)=\delta_{h,\mathbbm{1}_G}$.
	\item[$\circ$]{\it Antipode:}
	\be
		S(g \smo \delta_h) := g^{-1} \smo \delta_{g^{-1}h^{-1}g} \; .
	\ee
\end{itemize}
From the identification between the multiplication rule \eqref{mulDrin} of $\mD(G)$ and the gluing map \eqref{gluingCyl} of cylinder states, we deduce the correspondence
\be
	\label{corr2D}
	\mH_\mathbb{I} \ni	\cyli{g}{h} \longleftrightarrow \; (g \smo \delta_h)
	\in \mD(G) \; 
\ee
between cylinder (basis) states and Drinfel'd double (basis) elements. It follows that the elementary excitations or quasiparticles are labeled by the irreducible representations of $\mD(G)$ which provide the idempotents of the tube algebra \cite{ocneanu1993, ocneanu2001, DDR1}.

\subsection{Representation theory of $\mD(G)$}

The irreducible representations $\{\rho\}$ of $\mD(G)$ are labeled \cite{Koornwinder1999, Koornwinder1998} by a conjugacy class $C$ and an irreducible representation $R$ of the centralizer $Z_C$ of $C$ so that $\rho=(C,R)$. The elements of the conjugacy class $C$ are denoted $c_a$ and $c_1$ is chosen as representative. The centralizer $Z_C$ is then defined as the subgroup of elements commuting with the representative $c_1$ of $C$, i.e.
\be
Z_C = \{g \in G \, | \, gc_1 = c_1g \} \; .
\ee 
The elements of the quotient $Q_C \simeq G / Z_C$ are denoted $q_a$ and they satisfy the relation $c_a = q_a c_1 q_a^{-1}$. Finally, the matrix elements of the Drinfel'd double element $g \otimes \delta_h$ in the representation $\rho = (C,R)$ are given by
\be
	\boxed{
	D^{C,R}_{am,bn}(g \smo \delta_h) = \delta(h,c_a)\delta(c_a,gc_bg^{-1})D^R_{mn}(q_{a}^{-1}gq_b)
	}
\ee
where $m$ and $n$ are the matrix indices of the representation $R$ of $Z_C$, and the delta function $\delta(c_a,gc_bg^{-1})$ ensures that $q_a^{-1}gq_b$ belongs to $Z_C$. Thereafter, the more compact notation $D^{\rho}_{MN} \equiv D^{C,R}_{am,bn}$ is used, such that $M \equiv am$, $N \equiv bn$ and $\rho \equiv (C,R)$.  

The set $\{\rho\}$ of irreducible representations is complete and orthogonal. The completeness relation reads
\be
	\label{double_complete}
	\sum_{\rho}\sum_{M,N}d_{\rho}D^{\rho}_{MN}(g_1 \smo \delta_{h_1})\overline{D^{\rho}_{MN}(g_2 \smo \delta_{h_2})}
	= |G|\delta_{g_1,g_2}\delta_{h_1,h_2}
\ee
while the orthogonality is provided by
\be
	\label{double_ortho}
	\frac{1}{|G|}\sum_{g,h \in G}D^{\rho_1}_{M_1N_1}(g \smo \delta_h)\overline{D^{\rho_2}_{M_2N_2}(g \smo \delta_h)}
	= \frac{\delta_{\rho_1,\rho_2}}{d_{\rho_1}}\delta_{M_1,M_2}\delta_{N_1,N_2} \; ,
\ee
where $d_{\rho} = d_{C,R} = d_R \cdot |C|$ is the dimension of the representation $\rho$. Furthermore, thanks to the antipode map $S$, we can define the representation $\rho^\ast$ dual to the representation $\rho$ and the expression for the matrix elements is provided by 
\be
	D^{\rho^\ast}_{MN}(g \smo \delta_h) = D^\rho_{NM}(S(g \smo \delta_h)) = 	D^\rho_{NM}(g^{-1} \otimes \delta_{g^{-1}h^{-1}g})\; .
\ee
Thanks to the comultiplication, tensor product of representations can be defined such that
\be
	(D^{\rho_1}\otimes D^{\rho_2})(\Delta(g \otimes \delta_h)) = \sum_{x,y\in G \atop xy = h}(D^{\rho_1}\otimes D^{\rho_2})((g \otimes \delta_x)\otimes (g \otimes \delta_y)) \;.
\ee
Tensor products of representations can then be decomposed into irreducible representations according to the fusion rules $N^{\rho_3}_{\rho_1 \rho_2}$, i.e.
\be
	\label{double_fusRules}
	\rho_1 \otimes \rho_2 = \bigoplus_{\rho_3} N^{\rho_3}_{\rho_1 \rho_2}\, \rho_3 \;.
\ee
For notational convenience, we assume in the following that the fusion category of representations of the Drinfel'd double is multiplicity free, i.e. $N^{\rho_3}_{\rho_1\rho_2} \in \{0,1\}$.
Furthermore, the fact that the comultiplication is an algebra homomorphism implies the existence of a unitary map $\mathcal{C}^{\rho_1 \rho_2}: \bigoplus_{\rho_3 \in \rho_1 \otimes \rho_2}V_{\rho_3}\rightarrow V_{\rho_1} \otimes V_{\rho_2}$ which satisfies
\be
	D^{\rho_1}_{M_1N_1}\otimes D^{\rho_2}_{M_2N_2}(\Delta(g \otimes \delta_h)) = \sum_{\rho_3}\sum_{M_3N_3}
	\mathcal{C}^{\rho_1\rho_2\rho_3}_{M_1M_2M_3}\, D^{\rho_3}_{M_3N_3}(g \otimes \delta_h))\,\overline{\mathcal{C}^{\rho_1\rho_2\rho_3}_{N_1N_2N_3}} 
\ee
where $(M_1M_2)$ and $(\rho_3M_3)$ have to be understood as the indices of the matrix $\mC^{\rho_1 \rho_2}$. 
By analogy with the group case, these maps will be referred to as Clebsch-Gordan coefficients. In the following, it will be more convenient to work with the analogue of the Wigner $3 j m$-symbols, obtained by symmetrizing the Clebsch-Gordan coefficients
\be
	\Big({}^{\, \rho_1 \;\, \rho_2 \;\, \rho_3}_{M_1M_2M_3}\Big) :=
	\frac{1}{\sqrt{d_{\rho_3}}}\mC^{\rho_1 \rho_2 \rho_3^\ast}_{M_1M_2M_3} \; ,
\ee
which we will refer to as the $3\rho M$-symbols. The interwining map whose coefficients are given by the $3 \rho M$-symbols is denoted $\mathcal{I}^{\rho_1\rho_2\rho_3}$. The unitarity of $\mathcal{C}^{\rho_1 \rho_2}$ yields the orthogonality relation
\be
	\label{ortho_CG2D}
	\sum_{M_1,M_2}
	\Big({}^{\, \rho_1 \;\, \rho_2 \;\, \rho}_{M_1M_2M}\Big)
	\overline{\Big({}^{\, \rho_1 \;\, \rho_2 \;\, \rho'}_{M_1M_2M'}\Big)}
	= \frac{1}{d_{\rho}}\delta_{\rho,\rho'}\delta_{M,M'},
\ee
as well as the completeness relation
\be
	\label{complete_CG2D}
	\sum_{\rho}\sum_M d_\rho 
	\Big({}^{\, \rho_1 \;\, \rho_2 \;\, \rho}_{M_1M_2M}\Big)
	\overline{\Big({}^{\, \rho_1 \; \rho_2 \; \rho}_{N_1N_2M}\Big)}		
	 = \delta_{M_1,N_1}\delta_{M_2,N_2} \;.
\ee
Finally, it follows directly from the definition that the $3 \rho M$-symbols satisfy the invariance property \cite{DDR1}:
\be
	\label{invCG}
	\sum_{h_1,h_2}
	D^{\rho_1}_{M_1N_1}(g \smo \delta_{h_1})D^{\rho_2}_{M_2N_2}(g \smo \delta_{h_2})
	D^{\rho_3}_{M_3N_3}(g \smo \delta_{h_2^{-1}h_1^{-1}})\, 
	\Big({}^{\, \rho_1 \; \rho_2 \; \rho_3}_{N_1N_2N_3}\Big)
	= 	\Big({}^{\, \rho_1 \;\, \rho_2 \;\, \rho_3}_{M_1M_2M_3}\Big)\; .
\ee

\subsection{Excitation basis for (2+1)d topological phases}

We now have all the necessary ingredients to define our excitation basis for (2+1)d topological phases in terms of irreducible representations of the Drinfel'd double. This basis will be referred to as the {\it fusion basis} \cite{KKR, alagic2010estimating, DGTQFT, DDR1}. So far, we have defined basis states for the cylinder $\mathbb{I}$ and the two-torus $\mathbb{T}$ in terms of group holonomies. The correspondence \eqref{corr2D} provides the fusion basis states for the cylinder as the ``Fourier transform'' of the basis states \eqref{basisCyl2D}: 
\be
	\label{fusCyl2D}
	| \rho,MN \ra_\mathbb{I} = \frac{1}{|G|}\sum_{g,h \in G}\sqrt{d_{\rho}}
	D^{\rho}_{MN}(g \smo \delta_h)\cyli{g}{h} \; .
\ee
Such fusion basis states diagonalize the $\star$-multiplication:
\be
	| \rho_1,M_1N_1 \ra_\mathbb{I}\, \star | \rho_2,M_2N_2 \ra_\mathbb{I} 
	=\frac{\delta_{\rho_1,\rho_2}}{\sqrt{d_{\rho_1}}}\delta_{N_1,M_2}| \rho_1,M_1N_2\ra_\mathbb{I}	
\ee
which confirms that the fusion basis states \eqref{fusCyl2D} on the cylinder are the states of elementary quasi-excitations such that the conjugacy class $C$ labels fluxes while the representation $R$ labels charges. This means in particular that the irreducible representations trivialize the gluing operation presented in \eqref{starprod}. Indeed, with the fusion basis, the gluing boils down to a contraction of the states by summing over the corresponding magnetic indices. 
For instance, since the torus is nothing else than a cylinder with the pieces of its boundary identified, we deduce immediately that the fusion basis states for the torus read
\be
	\label{fus_twoTor}
	\boxed{
	| \rho \ra_\mathbb{T} = \frac{1}{|G|}\sum_{g,h \in G}
	\chi^{\rho}(g \smo \delta_h)\twotor{g}{h} = 
	\frac{1}{|G|}\sum_{h \in C \atop g \in Z_h}\chi^{R_h}(g) \twotor{g}{h}}
\ee
where $\chi^\rho$ denotes the character of the representation $\rho = (C,R)$, $Z_h = \{g \in G \, | \, gh=hg\}$ the centralizer of the group element $h$ and $\chi^{R_h}$ the character of the representation $R_h$ of $Z_h$ isomorphic to $R$. The vertices which once were at the boundary of the cylinder are identified to become a bulk vertex at which $\mathbb{A}_v$ acts. The group averaging induces the contraction of the magnetic indices which turns the representation matrix $D^\rho$ into the character $\chi^\rho$ defined as
\be
	\label{Fus2DTor}
	\chi^\rho(g \smo \delta_h) = \Theta_C(h)\delta_{gh,hg}\chi^R(q^{-1}_{\iota_C(h)}\, g \, q_{\iota_C(h)})
\ee 
where $\Theta_C({\sss \bullet})$ is the characteristic function of $C$ and $\iota_C({\sss \bullet})$ is a labeling function defined such that $\iota_C(c_a)=a$. Moreover, we have the following relation between characters of $R$ and $R_h$: $\chi^{R_h}(g) = \chi^R(q^{-1}_{\iota_C(h)}\, g \, q_{\iota_C(h)})$. Recall that the holonomy basis states, as represented in \eqref{fus_twoTor}, are already projected so that both the Gau{\ss} constraint and the zero-flux condition are imposed. However, these constraints are already encoded in the Fourier transform so that the fusion basis states for the two-torus can equivalently be rewritten $| \rho \ra_\mathbb{T} = \frac{1}{|G|}\sum_{h \in C \atop g \in Z_h}\chi^{R_h}(g)|g,h \ra_\mathbb{T}$. Furthermore, we recover with \eqref{fus_twoTor} the well-known result that the number of irreducible representations of $\mD(G)$ is equal to the ground state degeneracy on the torus, which counts quasi-excitation types \cite{Hu:2011pf}. 

In order to construct the fusion basis for arbitrary surfaces $\Sigma^p_\mathsf{g}$, we use the fact any punctured Riemann surface can be obtained by gluing together several copies of the thrice-punctured two-sphere $\mathbb{Y}$. A ``minimal'' thrice-punctured two-sphere can be discretized by a triangular face, on which $\mathbb{B}_f$ acts, with its three vertices identified. The corresponding basis states are labeled by two group holonomies associated to two independent non-contractible cycles and are represented by
\be
	\label{minThrice2D}
	\thrice{h_1}{h_2} 
	\longleftrightarrow
	\;\; \includegraphics[scale=1, valign=c]{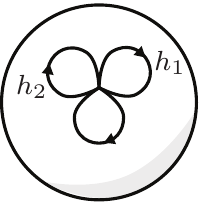}
\ee
where the identical dots represent identified vertices. In this example, the  holonomies $h_1$ and $h_2$ only account for magnetic degrees of freedom. The thrice-punctured sphere as discretized by \eqref{minThrice2D} is somewhat degenerate. In order to allow for point-like electric excitations associated to each one of the punctures, we need to choose a slightly more complicated discretization. This will support holonomies accounting for electric degrees of freedom, in addition to the ones accounting for magnetic degrees of freedom. Such discretization is obtained by gluing three outgoing cylinders to the minimal thrice-punctured sphere. The corresponding basis states span the Hilbert space $\mH_\mathbb{Y}$ and can be represented as 

\be
	\label{thrice2D}
	\mH_{\mathbb{Y}} = \Bigg\{
	\basisOne{g_3}{h_3}{g_1}{h_1}{g_2}{h_2} \Bigg\} 
	\q {\rm with} \q
	\basisOne{g_3}{h_3}{g_1}{h_1}{g_2}{h_2} \longleftrightarrow \includegraphics[scale=1, valign=c]{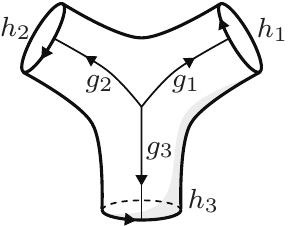} 
\ee
where the black-dotted vertex is now a bulk vertex at which the Gau{\ss} constraint is enforced.
It now remains to find the corresponding fusion basis. As suggested by the group representation, the basis states of $\mH_\mathbb{Y}$ can be obtained by gluing three outgoing cylinder basis states \eqref{basisCyl2D} to the thrice-punctured two-sphere state \eqref{minThrice2D}. The same happens for the fusion basis. The fusion basis states for the thrice-punctured sphere are defined as the gluing of three cylinder states $| \rho,MN \ra_\mathbb{I}$ via an intertwining map $\mathcal{I}^{\rho_1\rho_2\rho_3}$, i.e.
\begin{align}
	\label{fusThrice2D}
	| \{\rho_i,M_i\}_{i=1}^3\ra_\mathbb{Y} &={\rm tr}_{\{V^{\rho}\}}\big[ |\rho_1,M_1 \ra_\mathbb{I} \otimes 
	|\rho_2, M_2 \ra_\mathbb{I} \otimes | \rho_3,M_3 \ra_\mathbb{I} \otimes 
	\mathcal{I}^{\rho_1\rho_2\rho_3}
	\big] \\[0.5em] \nn
	&=
	\sum_{\{N_i\}_{i=1}^3} |\rho_1,M_1N_1 \ra_\mathbb{I} \otimes |\rho_2,M_2N_2 \ra_\mathbb{I} \otimes 
	|\rho_3,M_3N_3 \ra_\mathbb{I}\;
	\Big({}^{\, \rho_1 \; \rho_2 \; \rho_3}_{N_1N_2N_3}\Big) \\[-0.8em]\nn
	&=
	\frac{1}{|G|^3}\sum_{\{N_i\}_{i=1}^3}\sum_{\{g_i,h_i\}_{i=1}^3}\prod_{i=1}^3 \bigg(\sqrt{d_{\rho_i}}D^{\rho_i}_{M_iN_i}(g_i \smo \delta_{h_i})
	\cyli{g_i}{h_i}\bigg)
	\Big({}^{\, \rho_1 \; \rho_2 \; \rho_3}_{N_1N_2N_3}\Big) \; .
\end{align}
Thanks to the invariance property \eqref{invCG}, the intertwining map $\mathcal{I}^{\rho_1 \rho_2 \rho_3}$ ensures that the zero-flux condition on the closed surface of the thrice-punctured sphere is satisfied as well as the gauge invariance at the single bulk vertex. In \eqref{thrice2D}, the closed surface in question is the one represented by a triangle with its three vertices identified. Using the zero-flux conditions on the surface of each cylinder, we compute that imposing the zero-flux condition at the triangle boils down to the factor $\delta(g_1^{-1}h_1g_1g_2^{-1}h_2g_2g_3^{-1}h_3g_3,\mathbbm{1}_G)$. Let us work out how this zero-flux condition and the gauge invariance at the identified vertices are implicitly encoded in the $3\rho M$-symbols in \eqref{fusThrice2D}. Using the invariance property \eqref{invCG}, one has
\begin{align}
	\label{inv_Ex} \nn
	& \sum_{\{N_i\}_{i=1}^3}
	\bigg( \prod_{i=1}^3 D^{\rho_i}_{M_iN_i}(g_i \smo \delta_{h_i}) \bigg)
	\Big({}^{\, \rho_1 \; \rho_2 \; \rho_3}_{N_1N_2N_3}\Big) \\ \nn
	& \q =
	\sum_{k_1,k_2 \in G}
	\sum_{\{N_i\}_{i=1}^3}\sum_{\{O_i\}_{i=1}^3}
	\bigg( \prod_{i=1}^3 D^{\rho_i}_{M_iN_i}(g_i \smo \delta_{h_i}) \bigg) \\ \nn
	& \hspace{11em} \times 
	D^{\rho_1}_{N_1O_1}(g \smo \delta_{k_1})
	D^{\rho_2}_{N_2O_2}(g \smo \delta_{k_2})
	D^{\rho_3}_{N_3O_3}(g \smo \delta_{k_2^{-1}k_1^{-1}})
	\Big({}^{\, \rho_1 \; \rho_2 \; \rho_3}_{O_1O_2O_3}\Big) \\ \nn
	& \q = \sum_{k_1,k_2 \in G}\sum_{\{O_i\}_{i=1}^3}	
	\bigg( \prod_{i=1}^3D^{\rho_i}_{M_iO_i}(g_ig \smo \delta_{h_i})\bigg) 
	\delta_{k_1,g_1^{-1}h_1g_1}\delta_{k_2,g_2^{-1}h_2g_2}
	\delta_{k_2^{-1}k_1^{-1},g_3^{-1}h_3g_3}\Big({}^{\, \rho_1 \; \rho_2 \; \rho_3}_{O_1O_2O_3}\Big) \\
	& \q =
	\sum_{\{O_i\}_{i=1}^3}
	\bigg( \prod_{i=1}^3 D^{\rho_i}_{M_iO_i}(g_ig \smo \delta_{h_i}) \bigg)
	\delta(g_1^{-1}h_1g_1g_2^{-1}h_2g_2g_3^{-1}h_3g_3,\mathbbm{1}_G)
	\Big({}^{\, \rho_1 \; \rho_2 \; \rho_3}_{O_1 O_2 O_3}\Big)
\end{align}
where we used the defining property of the representations of the Drinfel'd double. By comparing the first and the last line of \eqref{inv_Ex}, we conclude that the $3 \rho M$-symbols implicitly encode the zero-flux condition as well as the gauge invariance. 
Note that we first introduced the comultiplication rule of the Drinfel'd double together with the corresponding tensor product and then use it to define states on the thrice-punctured two-sphere. Conversely, we could have first considered the gluing of three cylinder basis states as in \eqref{fusThrice2D} and derived which constraints needed to be imposed for this gluing to be consistent, from which we could have derived the comultiplication rule. 

Using the states \eqref{fusThrice2D}, we can construct a fusion basis for any Riemann surface $\Sigma_\mathsf{g}^p$. It suffices to decompose the surface $\Sigma_\mathsf{g}^p$ as a sewing of several copies of $\mathbb{Y}$, associate a state $|\{\rho_i,M_i\}_{i=1}^3 \ra_\mathbb{Y}$ to each copy of $\mathbb{Y}$ and contract them to each other following the decomposition pattern. Equivalently, we can associate an intertwining map $\mathcal{I}^{\rho_1\rho_2\rho_3}$ to each copy of $\mathbb{Y}$ and contract them via cylinder states $|\rho,MN \ra_\mathbb{I}$. For a given surface $\Sigma_\mathsf{g}^p$, and for a given pant decomposition $\{\mathbb{Y}\}$, a formal expression for the corresponding fusion basis states therefore reads
\be
	\boxed{
		\big| \{\rho\}\big\ra_{\Sigma_\mathsf{g}^p} = 
		{\rm tr}_{\{V^{\rho}\}}\Big[\bigotimes_\mathbb{I}|\rho \ra_\mathbb{I}
		\otimes \bigotimes^{ }_\mathbb{Y}\mathcal{I}^{\{\rho\}} \Big] 
	\; .}
\ee
The fusion basis is {\it orthogonal} and {\it complete} \cite{DDR1}. This follows directly from the orthogonality \eqref{double_ortho} and completeness \eqref{double_complete} of the representation matrices as well as the orthogonality \eqref{ortho_CG2D} and completeness \eqref{complete_CG2D} of the $3 \rho M$-symbols.

\vspace{2em}
\section{Three-dimensional generalization \label{sec:three}}
In this section, we generalize the previous construction to (3+1)d topological phases. We will extract the underlying algebraic structure from the gluing operation of the (3+1)d equivalents of the cylinder states. It will lead to a trialgebra which naturally extends the Drinfel'd double structure. Using the irreducible representations of this trialgebra, we will define a generalization of the fusion basis for (3+1)d topological phases with defect excitations.
\subsection{Three-cylinder algebra}
The previous definitions still hold in (3+1)d. In particular, the lattice Hamiltonian is the same as before \cite{Moradi:2014cfa, Wan:2014woa, 2017arXiv170404221L}, but defined with respect to a 3d lattice. This means that gauge invariance is still enforced at vertices and every flux going through a face associated with a contractible cycle is zero. For instance, the Hilbert space $\mH_{\mathbb{T}_3}$ of gauge invariant functionals on the space of flat connections on the three-torus $\mathbb{T}_3 = \mathbb{S}_1 \times \mathbb{S}_1 \times \mathbb{S}_1$ is given by
\be
	\nn
	\mH_{\mathbb{T}_3} = \big\{\frac{1}{|G|}\sum_{x \in G}|xgx^{-1},xhx^{-1},xkx^{-1} \ra_{\mathbb{T}_3} \, | \, 
	[g,h]=[g,k]=[h,k]=\mathbbm{1}_G \big\} =: \Bigg\{ \threeTor{g}{h}{k} \Bigg\}
\ee
where the discretization of the three-torus is composed of one cube, three faces on which $\mathbb{B}_f$ acts, three edges corresponding to the three non-contractible cycles, and one bulk vertex on which $\mathbb{A}_v$ acts. In $2$d, we obtained the cylinder (or twice-punctured two-sphere) $\mathbb{I}$ by cutting the two-torus along one direction. We proceed similarly in $3$d so as to obtain the topology $\mathbb{S}_1 \times \mathbb{S}_1 \times I$, with $I$ an interval. We will refer to the result of this cutting as the {\it three-cylinder} denoted by $\triCyl$. The boundary of the three-cylinder $\triCyl$ is the support of both point-like electric excitations and string-like magnetic excitations \cite{Moradi:2014cfa}. More precisely, the three-torus is bounded by two two-tori whose non-contractible cycles carry the magnetic excitations. Furthermore, each boundary torus carries one marked point, intersection of the non-contractible cycles, at which the Gau{\ss} constraint is relaxed so that the tori support both types of excitations.

Similarly to the $2$d cylinder, which is obtained by removing two disks from the two-sphere, we can obtain the three-cylinder by removing two linked solid two-tori from the three-sphere $\mathbb{S}_3$. This follows from the fact that the three-sphere can be obtained as the identification of two solid two-tori (this is the genus-one Heegaard splitting of the three-sphere). Removing one solid torus leaves us with the other solid torus which becomes $\triCyl$ after removing a second solid torus. Moreover, the three-cylinder is nothing but $\mathbb{I} \times \mathbb{S}_1$. This last remark will turn out to be very useful in the construction of the fusion basis for (3+1)d. We will restrict our analysis in this paper to the case where excitations are supported by torus boundaries.

The three-cylinder $\triCyl$ is therefore discretized by one cube, three faces on which $\mathbb{B}_f$ acts, five edges, and two boundary vertices at which the Gau{\ss} constraint is relaxed. The Hilbert space $\mH_{\triCyl}$ thus reads\footnote{Another way to visualize the three-cylinder is to think of it as a hollow two-torus so that the radial direction corresponds to the $g$-holonomy.}
\be
	\label{basisCyl3D}
	\mH_{\triCyl} = \Bigg\{ \threeCyl{g}{h}{k}  \Bigg\} \quad \text{with} \quad 
	\threeCyl{g}{h}{k} \longleftrightarrow \;\;\includegraphics[scale=1,valign=c]{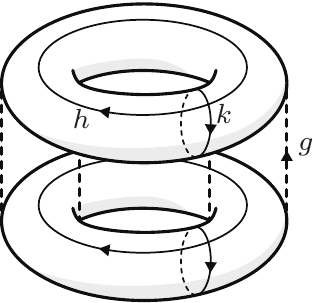}\; .
\ee 
Using these three-cylinder basis states, we can now repeat the gluing procedure in order to reveal the underlying structure of the excitations. The gluing follows the same rule as in two dimensions, however, during the identification step $\mathfrak{G}$, it is now necessary to identify the equatorial and the meridional non-contractible cycles of the boundary tori in addition to the marked points. Using our graphical representation for the states defined on the three-cylinder, the gluing reads 
\begin{align}
	\threeCyl{g_1}{h_1}{k_1} \star \threeCyl{g_2}{h_2}{k_2}
	&= (\mathbb{A} \circ \mathbb{B}) \triangleright \mathfrak{G}
	\Bigg( 	\threeCyl{g_1}{h_1}{k_1} , \threeCyl{g_2}{h_2}{k_2} \Bigg) \\
	&=
	\delta_{h_2,g_1^{-1}h_1g_1}\delta_{k_2,g_1^{-1}k_1g_1} \; 
	\doubleThreeCyl{g_1}{h_1}{k_1}{g_2}  \\
	&\sim 
	\delta_{h_2,g_1^{-1}h_1g_1}\delta_{k_2,g_1^{-1}k_1g_1}
	\threeCyl{g_1g_2}{h_1}{k_1}
\end{align}
where the last step repeatedly makes use of the equivalence relations \eqref{equivG} and \eqref{equivF}. We summarize this gluing operation as 
\be
	\label{gluingThreeCyl}
	\boxed{
	\threeCyl{g_1}{h_1}{k_1} \star \threeCyl{g_2}{h_2}{k_2} = 
	\delta_{k_2,g_1^{-1}k_1g_1}\delta_{h_2,g_1^{-1}h_1g_1}
	\threeCyl{g_1g_2}{h_1}{k_1}
	}
\ee
which is a (3+1)d generalization of Ocneanu's tube algebra. We will now describe how this gluing operation corresponds to the multiplication map of an algebraic structure which is a natural extension of the Drinfel'd double.

\section{Quantum triple $\mathcal{T}(G)$}
It is clear from \eqref{gluingThreeCyl} that if either $h=\mathbbm{1}_G$ or $k=\mathbbm{1}_G$, the algebra we are interested in reduces to the Drinfel'd double. We are therefore looking for an extension of the Drinfel'd double, denoted $\mT(G)$, and referred to as the {\it quantum triple} following the suggestion of \cite{Wang:2014oya}. Similarly to $\mD(G)$, which is obtained as the pairing between an algebra and its dual coalgebra with opposite comultiplication, $\mathcal{T}(G)$ can be thought as a trialgebra obtained from an algebra and two copies of its dual coalgebra. More generally, a trialgebra involves an associative algebra and two additional compatible algebraic structures such that the first one provides the bialgebra structure and the second one the trialgebra structure. 

Let us now propose the defining properties of the quantum triple $\mT(G)$. As a vector space, the quantum triple is isomorphic to 
\be
\mathcal{T}(G) \simeq \mathbb{C}[G] \otimes \mD\mathcal{F}(G) \subset \mathbb{C}[G] \otimes \mathcal{F}(G) \otimes \mathcal{F}(G)
\ee
where $\mD\mathcal{F}(G)$ denotes the abelian algebra of linear functions on $G \times G$ such that they have support on commuting holonomies only. A basis for $\mT(G)$ is therefore provided by $\{g \smo \delta_h \smo \delta_k \, | \, [h,k]=\mathbbm{1}_G\}_{g,h,k \in G}$. In the following, we will simply denote the basis elements by $g \smo \delta_h \smoc \delta_k \equiv \delta_{hk,kh}(g \smo \delta_h \smo \delta_k)$ where $\mathfrak{c}$ is there to remind of the commutation between the group variables $h$ and $k$. The quantum triple comes equipped with the maps:
\begin{itemize}
	\item[$\circ$]{\it Multiplication:}
	\be
	\label{mulTri}
	(g_1 \smo \delta_{h_1} \smoc \delta_{k_1})\star (g_2 \smo \delta_{h_2} \smoc \delta_{k_2}) := \delta_{k_2,g_1^{-1}k_1g_1}\delta_{h_2,g_1^{-1}h_1g_1}(g_1g_2 \smo \delta_{h_1} \smoc \delta_{k_1})
	\ee
	with corresponding unit element $\mathbbm{1}_{\mT(G)}=\sum_{h,k \in G} \mathbbm{1}_G \smo \delta_h \smoc \delta_k$. 
	\item[$\circ$]{\it Comultiplications:}
	\begin{align}
	\label{com3D}
	\Delta_{\rm I}(g \smo \delta_h \smoc \delta_k) := \sum_{x,y \in G \atop xy = h}
	(g \smo \delta_{x} \smoc \delta_{k}) \otimes (g \smo \delta_{y} \smoc \delta_{k}) \\ \nn
	\Delta_{\rm II}(g \smo \delta_h \smoc \delta_k) := \sum_{x,y \in G \atop xy = k}
	(g \smo \delta_{h} \smoc \delta_{x}) \otimes (g \smo \delta_{h} \smoc \delta_{y}) \; . 
	\end{align}
	\item[$\circ$]{\it Antipodes:}
	\begin{align}
	\label{anti3D}
	&S_{\rm I}(g \smo \delta_h \smoc \delta_k) := g^{-1} \smo \delta_{g^{-1}h^{-1}g} \smoc \delta_{g^{-1}kg} \\ \nn
	&S_{\rm II}(g \smo \delta_h \smoc \delta_k) := g^{-1} \smo \delta_{g^{-1}hg} \smoc \delta_{g^{-1}k^{-1}g} \; .
	\end{align}
\end{itemize}
As pointed out above, the quantum triple is defined such that if any of the coalgebras is `trivialized', i.e. reduced to the algebra of linear functions on the trivial subgroup $\{\mathbbm{1}_G\}$, we are left with a Drinfel'd double structure. Consequently, the defining maps (\ref{mulTri}, \ref{com3D}, \ref{anti3D}) satisfy relations very similar to the defining axioms of an Hopf algebra, and they satisfy them exactly when one of the coalgebras is trivialized. In particular, it follows straightforwardly from the compatibility conditions of the Drinfel'd double that the comultiplications $\Delta_{\rm I}$ and $\Delta_{\rm II}$ are algebra homomorphisms, i.e.
\begin{align}
\Delta_{\rm I,II} \big((g_1 \smo \delta_{h_1} \smoc \delta_{k_1}) \star (g_2 \smo \delta_{h_2} \smoc \delta_{k_2}) \big) &=
\Delta_{\rm I,II}(g_1 \smo \delta_{h_1} \smoc \delta_{k_1}) \star \Delta_{\rm I,II}(g_2 \smo \delta_{h_2} \smoc \delta_{k_2}) \; ,
\end{align}
while the antipodes $S_{\rm I}$ and $S_{\rm II}$ are algebra antihomomorphisms, i.e.
\begin{align}
S_{\rm I,II} \big((g_1 \smo \delta_{h_1} \smoc \delta_{k_1}) \star (g_2 \smo \delta_{h_2} \smoc \delta_{k_2}) \big) &=
S_{\rm I,II}(g_2 \smo \delta_{h_2} \smoc \delta_{k_2}) \star S_{\rm I,II}(g_1 \smo \delta_{h_1} \smoc \delta_{k_1})  \; .
\end{align}
The fundamental difference between the quantum triple $\mT(G)$ and the Drinfel'd double $\mD(G)$ is naturally the existence of two comultiplications. These two maps naturally yield two notions of tensor product, and {\it a fortiori}, to two different sets of fusion rules. Moreover, because of the existence of antihomomorphic antipode maps $S_{\rm I}$ and $S_{\rm II}$, it is possible to define dual representations with respect to either $\Delta_{\rm I}$ or $\Delta_{\rm II}$. Accordingly, we can also define two notions of trivial representations. Moreover, one representation is trivial with respect to both $\Delta_{\rm I}$ and $\Delta_{\rm II}$, and it is defined in terms of the counit $\epsilon(g \smo \delta_h \smoc \delta_k) = \delta_{h,\mathbbm{1}_G}\delta_{k,\mathbbm{1}_G}$. 

From the identification between the multiplication rule \eqref{mulTri} of $\mT(G)$ and the gluing map \eqref{gluingThreeCyl} of three-cylinder states, we deduce the correspondence
\be
\label{corr3D}
\mH_{\mathbb{I}_3} \owns	\threeCyl{g}{h}{k} \longleftrightarrow \; (g \smo \delta_h \smoc \delta_k)
\in \mT(G) \; 
\ee
between three-cylinder (basis) states and quantum triple (basis) elements.

\section{Representation theory of $\mT(G)$}
So far the properties of the quantum triple $\mT(G)$ have followed closely the ones of the Drinfel'd double $\mD(G)$. The same is true for the representation theory of $\mT(G)$ which is a natural extension of the one of $\mD(G)$. 

Recall that the irreducible representations $\{\rho\}$ of $\mD(G)$ are labeled by a conjugacy class $C(G)$ of the full group $G$, and an irreducible representation $R(Z_C)$ of the centralizer $Z_C$ of $C$, so that $\rho=(C(G),R(Z_C))$. It turns out that the irreducible representations $\{\wp\}$ of $\mT(G)$ are labeled by a conjugacy class $C(G)$ of $G$, a conjugacy class $D(Z_C)$ of the centralizer $Z_C$ of $C$, and an irreducible representation $R(Z_D)$ of the centralizer $Z_D$ of $D$ so that $\wp=(C(G),D(Z_C),R(Z_D))$. Naturally, when $C(G)=\{\mathbbm{1}_G\}$, the irreducible representations $\{\wp\}$ of $\mT(G)$ reduce to the representations $\{\rho\}$ of $\mD(G)$. 

The elements of the conjugacy class $C(G)$ are denoted $c_a$ and $c_1$ is chosen as representative so that the centralizer $Z_C$ is defined as $\{g \in G \, | \, gc_1=c_1g\}$. The elements of the quotient $P_C \simeq G / Z_C$ are denoted $p_a$ and they satisfy the relation $c_a = p_a c_1 p_a^{-1}$. The elements of the conjugacy class $D(Z_C)$ are denoted $d_\alpha$ and $d_1$ is chosen as representative so that the centralizer $Z_D$ is defined as $\{g \in Z_C \, | \, gd_1=d_1g\}$. The elements of the quotient $Q_D \simeq Z_C / Z_D$ are denoted $q_\alpha$ and they satisfy the relation $d_\alpha = q_\alpha d_1 q_\alpha^{-1}$. Finally, the matrix elements of the quantum triple element $g \smo \delta_h \smoc \delta_k$ in the representation $\wp = (C(G),D(Z_C),R(Z_D))$ are given by
\begin{empheq}[box=\fbox]{align} \nn
	D^{C,D,R}_{a\alpha m,b \beta n}(g \smo \delta_h \smo \delta_k) &= \delta(k,c_a)\delta(c_a,gc_bg^{-1})\delta(hk,kh) \\
	&\times
	\delta(p_a^{-1}hp_a,d_\alpha)
	\delta(d_\alpha,p_a^{-1}gp_bd_\beta p_b^{-1}g^{-1}p_a)D^R_{mn}(q_{\alpha}^{-1}p_a^{-1}gp_bq_\beta)
\label{TripleFT}
\end{empheq}
where $\delta(c_a,gc_bg^{-1})$ ensures that $p_a^{-1}gp_b$ belongs to $Z_C$ and $\delta(d_\alpha,p_a^{-1}gp_bd_\beta p_b^{-1}g^{-1}p_a)$ ensures that $q_{\alpha}^{-1}p_a^{-1}gp_bq_\beta$ belongs to $Z_D$. It is easy to check explicitly, using the definition of $c_a$ and $d_\alpha$, that $q_{\alpha}^{-1}p_a^{-1}gp_bq_\beta$ commutes with both $c_1$ and $d_1$. Furthermore, since $hk=kh$, we have $p_a^{-1}hp_a \in Z_C$. Note that we dropped the label $\mathfrak{c}$ in the tensor product since the commutation of the group variables $h$ and $k$ is now encoded in the definition of the representations. Thereafter, the more compact notation $D^{\wp}_{MN} \equiv D^{C,D,R}_{a\alpha m,b \beta n}$ is used, such that $M \equiv a\alpha m$, $N \equiv b \beta n$ and $\wp \equiv (C,D,R)$. Interestingly, an alternative basis can be  defined where the role of the group variables $h$ and $k$ is switched, i.e.
\begin{align}\nn
	D^{D,C,R}_{a\alpha m,b \beta n}(g \smo \delta_h \smo \delta_k) &= \delta(h,c_a)\delta(c_a,gc_bg^{-1})\delta(hk,kh) \\
	&\times
	\delta(p_a^{-1}kp_a,d_\alpha)
	\delta(d_\alpha,p_a^{-1}gp_bd_\beta p_b^{-1}g^{-1}p_a)D^R_{mn}(q_{\alpha}^{-1}p_a^{-1}gp_bq_\beta) \; .
	\label{TripleFT2}
\end{align}
Both the bases defined above are compatible with the algebraic structure. Using for instance def.~\eqref{TripleFT}, we can show that these representations are algebra homomorphisms (see app.~\ref{app_defprop} for proof), i.e.
\begin{align}
\sum_N D^\wp_{MN}(g_1 \smo \delta_{h_1} \smo \delta_{k_1})
D^\wp_{NO}(g_2 \smo \delta_{h_2} \smo \delta_{k_2}) = 
D^\wp_{MO}((g_1 \smo \delta_{h_1} \smo \delta_{k_1}) \star
(g_2 \smo \delta_{h_2} \smo \delta_{k_2})) \; .
\end{align}
Furthermore, the set $\{\wp\}$ of irreducible representations is complete and orthogonal. 
The completeness relation reads
\be
\label{triple_complete}
\sum_{\wp}\sum_{M,N}d_{\wp}D^{\wp}_{MN}(g_1 \smo \delta_{h_1} \smo \delta_{k_1})\overline{D^{\wp}_{MN}(g_2 \smo \delta_{h_2} \smo \delta_{k_2})}
= |G|\delta_{g_1,g_2}\delta_{h_1,h_2}\delta_{k_1,k_2}\delta_{h_1k_1,k_1h_1}
\ee
while the orthogonality is provided by ({\it cf} app.~\ref{app_ortho} for proof)
\be
\label{triple_ortho}
\frac{1}{|G|}\sum_{g,h,k \in G}
D^{\wp_1}_{M_1N_1}(g \smo \delta_h \smo \delta_k)\overline{D^{\wp_2}_{M_2N_2}(g \smo \delta_h \smo \delta_k)}
= \frac{\delta_{\wp_1,\wp_2}}{d_{\wp_1}}\delta_{M_1,M_2}\delta_{N_1,N_2} 
\ee
where $d_{\wp} = d_{C,D,R} = d_R \cdot |C| \cdot |D|$ is the dimension of the representation $\wp$. 

As a trialgebra, the quantum triple comes equipped with two compatible coalgebraic structures. In particular, this means there exists two different comultiplication maps which can be used to define two different kinds of tensor product of irreducible representations. Therefore, we can define two sets of fusion rules, associated with each one of the comultiplication maps $\Delta_{\rm I}$ and $\Delta_{\rm II}$. However, given the choice of basis for the representations \eqref{TripleFT} or \eqref{TripleFT2}, only one coalgebraic structure is compatible. Because of the obvious symmetry between $\Delta_{\rm I}$ and $\Delta_{\rm II}$, it is enough to focus on $\Delta_{\rm I}$. This is the choice compatible with the basis \eqref{TripleFT}. In that case, the tensor product reads
\begin{align} \nn
&(D^{\wp_1}\otimes_{\rm I} D^{\wp_2})(\Delta_{\rm I}(g \smo \delta_h \smo \delta_k)) = \sum_{x,y \in G \atop xy = h}
(D^{\wp_1}\otimes_{\rm I} D^{\wp_2})((g \smo \delta_{x} \smo \delta_{k})\otimes (g \smo \delta_{y} \smo \delta_{k})) 
\end{align}
which can be decomposed into irreducible representations according to the fusion rules ${}_{\rm I}N^{\wp_3}_{\wp_1 \wp_2}$, i.e.
\be
\wp_1 \otimes_{\rm I} \wp_2 = \bigoplus_{\wp_3} {}_{\rm I}N^{\wp_3}_{\wp_1 \wp_2}\, \wp_3 \; .
\ee
Again, we assume for notational convenience the multiplicity freeness of the corresponding fusion category, i.e. $N^{\wp_3}_{\wp_1 \wp_2} \in \{0,1\}$.
These fusion rules can be explicitly obtained in terms of the characters $\chi^\wp$. For instance, we have
\begin{align}
{}_{\rm I}N^{\wp_3}_{\wp_1 \wp_2} &= \frac{1}{|G|}
\sum_{g,h,k \in G}{\rm tr}[D^{\wp_1} \otimes_{\rm I} D^{\wp_2}](\Delta_{\rm I}(g \smo \delta_h \smo \delta_k))\overline{
	\chi^{\wp_3}(g \smo \delta_h \smo \delta_k)} \\
&= \frac{1}{|G|}
\sum_{g,k \in G}\sum_{h_1,h_2 \in G}\chi^{\wp_1}(g \smo \delta_{h_1} \smo \delta_k) \chi^{\wp_2}(g \smo \delta_{h_1^{-1}h_2} \smo \delta_k))\overline{
	\chi^{\wp_3}(g \smo \delta_{h_2} \smo \delta_k)} 
\; .
\end{align}
Several important remarks can be drawn from this last equation. Firstly, the fusion of representations with respect to the comultiplication $\Delta_{\rm I}$ vanish if the conjugacy class $C$ is not the same for $\wp_1$, $\wp_2$ and $\wp_3$. Secondly, the fusion rules for the quantum triple effectively boil down to the ones of $\mD(G)$, but they are parametrized by the choice of conjugacy class $C$. More precisely, in the case of the comultiplication $\Delta_{\rm I}$ and for a given conjugacy class $C$, the fusion rules of $\mT(G)$ boil down to the ones of the Drinfel'd double $\mD(Z_C)$ for the subgroup $Z_C$. It therefore suggests that for the fusion category $\text{Rep}[\mathcal{T}(G)]$ formed by the representations of $\mathcal{T}(G)$ the following grading holds:\footnote{This isomorphism is not true at the level of the vector spaces since it would require the $g$-holonomy and the $k$-holonomy to commute in the definition of the basis elements.}
\be
\text{Rep}[\mT(G)] \simeq \bigoplus_C \text{Rep}[\mD(Z_C)] \; . 
\ee
This obviously reminds of dimensional reduction which consists in expressing (3+1)d topological orders as a sum of (2+1)d topological orders via a compactification of one of the spatial directions. The conjugacy class $C$ is associated to such compactified direction. The fact that the quantum triple is equipped with two comultiplication maps only translates the fact that we can think of either the $h$-holonomy or the $k$-holonomy of the three-cylinder as being along the compactified direction. This also determines a choice of basis for the representation matrices.

Thanks to the antipode map $S_{\rm I}$, we can define the representations $\wp^{\ast}$ dual to the representation $\wp$ with respect to the set of fusion rules defined above. The corresponding expressions for the matrix elements are provided by 
\begin{align}
\label{def_dualrep}
&D^{\wp^{\ast}}_{MN}(g \smo \delta_h \smo \delta_k) = D^\wp_{NM}(S_{\rm I}(g \smo \delta_h \smo \delta_k)) = 	D^\wp_{NM}(g^{-1} \smo \delta_{g^{-1}h^{-1}g} \smo \delta_{g^{-1}kg}) 
\; .
\end{align}
Furthermore, since the comultiplication $\Delta_{\rm I}$ is an algebra homomorphism with respect to $\star$, we can define unitary maps 
\be
{}_{\rm I}\mathcal{C}^{\wp_1 \wp_2}: \bigoplus_{\wp_3 \in \wp_1 \otimes \wp_2}V_{\wp_3}\rightarrow V_{\wp_1} \otimes_{\rm I} V_{\wp_2}
\ee
which satisfy
\begin{align} \nn
&D^{\wp_1}_{M_1N_1}\otimes_{\rm I} D^{\wp_2}_{M_2N_2}(\Delta_{\rm I}(g \smo \delta_h \smo \delta_k)) = \sum_{\wp_3}\sum_{M_3,N_3}
{}_{\rm I}\mathcal{C}^{\wp_1\wp_2\wp_3}_{M_1M_2M_3}\, D^{\wp_3}_{M_3N_3}(g \smo \delta_h \smo \delta_k)\,{}_{\rm I}\overline{\mathcal{C}^{\wp_1\wp_2\wp_3}_{N_1N_2N_3}} 
\end{align}
where $(M_1M_2)$ and $(\wp_3M_3)$ have to be understood as the indices of the matrix ${}_{\rm I}\mC^{\wp_1 \wp_2}$. As before, we define the more symmetric symbols
\be
\Big({}^{\, \wp_1 \;\, \wp_2 \;\, \wp_3}_{M_1M_2M_3}\Big)_{\rm I} :=
\frac{1}{\sqrt{d_{\wp_3}}}{}_{\rm I}\mC^{\wp_1 \wp_2 \wp_3^{\ast}}_{M_1M_2M_3} 
\ee
which we will refer to as the $3\wp M$-symbols. The intertwining maps whose coefficients are given by the $3 \wp M$-symbols are denoted $\mathcal{I}_{\rm I}^{\wp_1\wp_2\wp_3}$. The unitarity of ${}_{\rm I}\mathcal{C}^{\wp_1 \wp_2}$ yields the orthogonality relation
\be
\label{ortho_CG3D}
\sum_{M_1,M_2}
\Big({}^{\, \wp_1 \;\, \wp_2 \;\, \wp}_{M_1M_2M}\Big)_{\rm I}
\overline{\Big({}^{\, \wp_1 \;\, \wp_2 \;\, \wp'}_{M_1M_2M'}\Big)_{\rm I}}
= \frac{1}{d_{\wp}}\delta_{\wp,\wp'}\delta_{M,M'},
\ee
as well as the completeness relation
\be
\label{complete_CG3D}
\sum_{\wp}\sum_M d_\wp 
\Big({}^{\, \wp_1 \;\, \wp_2 \;\, \wp}_{M_1M_2M}\Big)_{\rm I}
\overline{\Big({}^{\, \wp_1 \; \wp_2 \; \wp}_{N_1N_2M}\Big)_{\rm I}}		
= \delta_{M_1,N_1}\delta_{M_2,N_2} \;.
\ee
Finally, it follows directly from the definition that the $3 \wp M$-symbols satisfy the invariance property
\begin{align}
\label{invCG3} 
\Big({}^{\, \wp_1 \;\, \wp_2 \;\, \wp_3}_{M_1M_2M_3}\Big)_{\rm I}  &= 
\sum_{h_1,h_2,k}
D^{\wp_1}_{M_1N_1}(g \smo \delta_{h_1} \smo \delta_{k})
D^{\wp_2}_{M_2N_2}(g \smo \delta_{h_2} \smo \delta_{k})
D^{\wp_3}_{M_3N_3}(g \smo \delta_{h_2^{-1}h_1^{-1}} \smo \delta_{k}) 	\Big({}^{\, \wp_1 \; \wp_2 \; \wp_3}_{N_1N_2N_3}\Big)_{\rm I} 
\end{align}
which is proven in app.~\eqref{app_invCG}. In the following, we will use these intertwining maps to define a generalization of the fusion basis to (3+1)d. The existence of two sets of fusions rules, and their corresponding intertwining maps, suggests that there are two geometrically different ways of fusing torus-excitations. Nevertheless, we restrict our attention to one type of fusion rules only. This means that all the tensor products will be defined with respect to the same comultiplication, namely $\Delta_{\rm I}$.

\section{Excitation basis for (3+1)d topological phases}

As for the (2+1)d case, the representation theory of the quantum triple $\mT(G)$ provides us with a natural way of defining the so-called fusion basis for excited states. The (2+1)d construction relied on the fact that any Riemann surface $\Sigma^p_\mathsf{g}$ can be decomposed into thrice-punctured two-sphere $\mathbb{Y}$. Such a general statement does not exist for $3$d manifolds. Nevertheless, we have the following result: Any three-manifolds of the form $\Sigma^p_\mathsf{g} \times \mathbb{S}_1$ can be obtained by gluing several copies of the manifold $\mathbb{Y} \times \mathbb{S}_1$. As we shall see, this is reminiscent of the fact that the three-cylinder $\mathbb{I}_3$ can be obtained as $\mathbb{I} \times \mathbb{S}_1$. The manifold $\mathbb{Y} \times \mathbb{S}_1$ is bounded by three copies of the two-torus $\mathbb{T}_2$. Therefore, by considering manifolds of the form $\Sigma^p_\mathsf{g} \times \mathbb{S}_1$, we are constructing a basis for topological phases with defect excitations located at boundary two-tori.

To construct the generalization of the fusion basis to (3+1)d topological phases, we will follow step by step the previous construction. Everytime we considered a surface $\Sigma$ in (2+1)d, we will now look at the manifold $\Sigma \times \mathbb{S}_1$. In other words, we could first define the fusion basis for $\Sigma$ and then take the direct product with $\mathbb{S}_1$, hence lifting the Drinfel'd double elements to quantum triple elements. Naturally the basis states will now be labeled by irreducible representations of the quantum triple. The resulting basis will also be referred to as the \emph{fusion basis}. 

So far, we have defined basis states for the three-cylinder $\mathbb{I}_3$ and the three-torus $\mathbb{T}_3$ in terms of group holonomies. The correspondence \eqref{corr3D} provides the fusion basis states for the three-cylinder as the `Fourier transform' of the basis states \eqref{basisCyl3D}: 
\be
\label{fusCyl3D}
| \wp,MN \ra_{\mathbb{I}_3} = \frac{1}{|G|}\sum_{g,h,k \in G}\sqrt{d_{\wp}}
D^{\wp}_{MN}(g \smo \delta_h \smo \delta_k)\threeCyl{g}{h}{k} \; .
\ee
By construction, these fusion basis states diagonalize the $\star$-multiplication:
\be
| \wp_1,M_1N_1 \ra_{\mathbb{I}_3} \, \star | \wp_2,M_2N_2 \ra_{\mathbb{I}_3} 
=\frac{\delta_{\wp_1,\wp_2}}{\sqrt{d_{\wp_1}}}\delta_{N_1,M_2}| \wp_1,M_1N_2\ra_{\mathbb{I}_3}	
\ee
which confirms that the fusion basis states \eqref{fusCyl3D} on the three-cylinder are the states of elementary quasi-excitations. Analogously to (2+1)d, the conjugacy classes $C$ and $D$ are associated with fluxes and the representation $R$ with charges. The obvious difference between (2+1)d and (3+1)d is therefore that the quasi-excitations carry two flux labels. Note however that these two labels are independent only in the case where the group $G$ is abelian.

In particular, since the three-torus is nothing else than a three-cylinder with the pieces of its boundary identified, we deduce immediately that the fusion basis states for the three-torus read \cite{Wan:2014woa}\\[0.5em]
\be
\label{fus_threeTor}
\boxed{
	| \wp \ra_{\mathbb{T}_3} = \frac{1}{|G|}\sum_{g,h,k \in G}
	\chi^{\wp}(g \smo \delta_h \smo \delta_k)\threeTor{g}{h}{k} = 
	\frac{1}{|G|}\sum_{k \in C, h \in D(Z_k) \atop g \in Z_{h,k}}\chi^{R_{h,k}}(g) |g,h,k\rangle_{\mathbb{T}_3}}
\vspace{1em}
\ee
where $Z_{h,k} = \{g \in G \, | \, gh=hg\, , \, gk=kg\}$ denotes the centralizer of both the group elements $h$ and $k$ and $\chi^{R_{h,k}}$ the character of the representation $R_{h,k}$ of $Z_{h,k}$ isomorphic to $R$. The two vertices which were located at the boundary of the three-cylinder are now identified so that the three-torus has a single bulk vertex at which $\mathbb{A}_v$ acts. The group averaging induces the contraction of the magnetic indices $M$ and $N$ which turns the representation matrix $D^\wp$ into the character $\chi^\wp$ defined as
\begin{align} \nn
\chi^\wp(g \smo \delta_h \smo \delta_k) &= \Theta_C(k)\Theta_D(p_{\iota_C(k)}^{-1}hp_{\iota_C(k)})\delta_{gh,hg}\delta_{gk,kg}\delta_{hk,kh} \\
& \times \chi^R \big (q^{-1}_{\iota_D(p_{\iota_C(k)}^{-1}hp_{\iota_C(k)})}p_{\iota_C(k)}^{-1}\, g \, p_{\iota_C(k)}q_{\iota_D(p_{\iota_C(k)}^{-1}hp_{\iota_C(k)})} \big)
\end{align}
where $\Theta_C({\sss \bullet})$ and $\Theta_D({\sss \bullet})$ denote the characteristic functions of the conjugacy classes $C$ and $D$, respectively, $\iota_C({\sss \bullet})$ and $\iota_D({\sss \bullet})$ are labeling functions for $C$ and $D$ defined such that $\iota_C(c_a)=a$ and $\iota_D(d_\alpha)=\alpha$, respectively. Moreover, we can now write explicitly the relation between the characters of $R_{h,k}$ and $R$: $\chi^{R_{h,k}}(g) =  \chi^R \big (q^{-1}_{\iota_D(p_{\iota_C(k)}^{-1}hp_{\iota_C(k)})}p_{\iota_C(k)}^{-1}\, g \, p_{\iota_C(k)}q_{\iota_D(p_{\iota_C(k)}^{-1}hp_{\iota_C(k)})} \big)$.

As in the (2+1)d, the ground-states on the three-torus are in one-to-one correspondence with the quasi-excitations defined on the three-cylinder. Remark that, as for the representation matrices, this is not the only basis possible. As a matter of fact we can define six equivalent bases which correspond to the six different ways to `order' the variables $g$, $h$ and $k$.

As explained above, in order to construct the fusion basis associated to surfaces of the form $\Sigma \times \mathbb{S}_1$, we need first to consider the fusion basis states for the manifold $\mathbb{Y} \times \mathbb{S}_1$. Knowing that the `minimal' thrice-punctured two-sphere can be discretized by a triangular face whose vertices are identified, we deduce that $\mathbb{Y} \times \mathbb{S}_1$ can be minimaly discretized by a triangular prism whose six vertices are identified. The basis states associated to such a discretization are  labeled by three group holonomies corresponding to the three non-contractible cycles and are represented by \\[0.7em]
\be
\vspace{0.5em}
\label{minThrice3D}
\thriceThree{h_1}{h_2}{k} \; ,
\ee
where the dots represent identified vertices. 

Recall furthermore that the manifold $\mathbb{Y} \times \mathbb{S}_1$ is bounded by three two-tori. Exactly as in the (2+1)d case, the discretization \eqref{minThrice3D} is somewhat degenerate so that we would like to consider a slightly more complicated discretization which allows to associate a set of group holonomies $\{g,h,k \in G \, | \, hk=kh\}$ with each of one of these tori. This discretization is obtained by gluing three three-cylinder $\mathbb{I}_3$ to each one of the square faces of \eqref{minThrice3D}. The holonomies $\{h\}$ and $\{k\}$ then account for string-like magnetic degrees of freedom while the holonomies $\{g\}$ account for point-like electric degrees of freedom. Note however that because of the geometry of the three-cylinder states, such a gluing can be performed in two different ways, or more precisely along two different orientations. Either we decide to associate the $\{k\}$-holonomies to the $\mathbb{S}_1$ direction, or the $\{h\}$-holonomies. In terms of representations, this determines the choice of comultiplication map. Because of the symmetry between the two coalgebraic structures, both possibilities are equivalent, however, for consistency requirements all the gluing must be performed according to the same orientation so that we obtain a topology of the form $\Sigma \times \mathbb{S}_1$. In the following, we will choose the orientation consistent with the graphical representation presented above so that the $k$-holonomy always refers the $\mathbb{S}_1$ direction. The comultiplication compatible with this choice is $\Delta_{\rm I}$. As such, the states defined above provide a geometrical interpretation of the fusion rules ${}_{\rm I}N$. In (2+1)d, the fusion of excitations can be imagined as replacing two punctures by a single one containing the original ones. The fusion of defects in (3+1)d boils down to the (2+1)d picture with an additional direct product with the circle.

Because of the zero-flux condition located at the triangle of the discretization \eqref{minThrice3D}, we know that, after gluing of the three $\mathbb{I}_3$ states, there will be the same constraint between $\{g\}$ and $\{h\}$-holonomies as in the (2+1)d case. There will be a further constraint which identifies the holonomies $\{g^{-1}kg\}$. This last constraint might seem surprising. It is actually reminiscent of the fact that we are working with a manifold of the form $\Sigma \times \mathbb{S}_1$ and therefore, there is only one independent holonomy in the $\mathbb{S}_1$ direction. This also justifies why we are working with comultiplication maps of the form \eqref{com3D}. Indeed, we are dealing with fusion rules which ensure that the conjugacy class associated with one of the spatial directions always remain the same. This conjugacy class is the one associated with the $\mathbb{S}_1$ direction of the manifold under consideration.

It now remains to find the fusion basis states defined on the manifold $\mathbb{Y} \times \mathbb{S}_1$. The construction above suggests that one can obtain the basis states of $\mH_{\mathbb{Y} \times \mathbb{S}_1}$ by gluing three three-cylinder fusion basis states $| \wp,MN \ra_{\mathbb{I}_3}$ via an intertwining map $\mathcal{I}^{\wp_1\wp_2\wp_3}_{\rm I}$, i.e.
\begin{align}
\label{fusThrice3D}
| \{\wp_i,M_i\}_{i=1}^3\ra_\mathbb{Y} &={\rm tr}_{\{V^{\wp}\}}\big[ |\wp_1, M_1 \ra_\mathbb{I} \otimes_{\rm I} 
|\wp_2, M_2 \ra_\mathbb{I} \otimes_{\rm I} | \wp_3, M_3 \ra_\mathbb{I} \otimes \mathcal{I}_{\rm I}^{\wp_1 \wp_2 \wp_3}\big] \\[0.6em] \nn
&=
\sum_{\{N_i\}_{i=1}^3} |\wp_1,M_1N_1 \ra_\mathbb{I} \otimes_{\rm I} |\wp_2,M_2N_2 \ra_\mathbb{I} \otimes_{\rm I} 
|\wp_3,M_3N_3 \ra_\mathbb{I}\;
\Big({}^{\, \wp_1 \; \wp_2 \; \wp_3}_{N_1N_2N_3}\Big)_{\rm I} \\[-0.8em] \nn
&=
\frac{1}{|G|^3}\sum_{\{M_i\}_{i=1}^3}\sum_{\{g_i,h_i,k_i\}}\prod_{i=1}^3 \bigg(\sqrt{d_{\wp_i}}D^{\wp_i}_{M_iN_i}(g_i \smo \delta_{h_i} \smo \delta_{k_i})
\threeCyl{g_i}{h_i}{k_i}\bigg)
\Big({}^{\, \wp_1 \; \wp_2 \; \wp_3}_{N_1N_2N_3}\Big)_{\rm I} \; .
\end{align}
Thanks to the invariance property \eqref{invCG3} of the intertwining map $\mathcal{I}_{\rm I}^{\wp_1 \wp_2 \wp_3}$ and following exactly the same steps as in \eqref{inv_Ex}, one has
\begin{align}
\label{inv_Ex2} \nn
& \sum_{\{N_i\}_{i=1}^3}
\bigg( \prod_{i=1}^3 D^{\wp_i}_{M_iN_i}(g_i \smo \delta_{h_i} \smo \delta_{k_i}) \bigg)
\Big({}^{\, \wp_1 \; \wp_2 \; \wp_3}_{N_1N_2N_3}\Big)_{\rm I} \\ \nn
& \q =
\sum_{p_1,p_2,k \in G}
\sum_{\{N_i\}_{i=1}^3}\sum_{\{O_i\}_{i=1}^3}
\bigg( \prod_{i=1}^3 D^{\wp_i}_{M_iN_i}(g_i \smo \delta_{h_i} \smo \delta_{k_i}) \bigg) \\ \nn
& \hspace{6em} \times 
D^{\wp_1}_{N_1O_1}(g \smo \delta_{p_1} \smo \delta_k)
D^{\wp_2}_{N_2O_2}(g \smo \delta_{p_2} \smo \delta_k)
D^{\wp_3}_{N_3O_3}(g \smo \delta_{p_2^{-1}p_1^{-1}} \smo \delta_k)
\Big({}^{\wp_1 \, \wp_2 \, \wp_3}_{O_1O_2O_3}\Big)_{\rm I} \\ \nn
& \q = \sum_{p_1,p_2,k \in G}\sum_{\{O_i\}_{i=1}^3}	
\bigg( \prod_{i=1}^3D^{\wp_i}_{M_iO_i}(g_ig \smo \delta_{h_i} \smo \delta_{k_i})\bigg) \\ \nn &\hspace{6em} \times 
\delta_{p_1,g_1^{-1}h_1g_1}\delta_{p_2,g_2^{-1}h_2g_2}
\delta_{p_2^{-1}p_1^{-1},g_3^{-1}h_3g_3}
\delta_{k,g_1^{-1}k_1g_1}\delta_{k,g_2^{-1}k_2g_2}\delta_{k,g_3^{-1}k_3g_3}
\Big({}^{\wp_1 \, \wp_2 \, \wp_3}_{O_1O_2O_3}\Big)_{\rm I} \\ \nn
& \q =
\sum_{\{O_i\}_{i=1}^3}
\bigg( \prod_{i=1}^3 D^{\wp_i}_{M_iO_i}(g_ig \smo \delta_{h_i} \smo \delta_{k_i}) \bigg)
\delta(g_1^{-1}h_1g_1g_2^{-1}h_2g_2g_3^{-1}h_3g_3,\mathbbm{1}_G)
\\  &\hspace{6em} \times \delta_{g_1^{-1}k_1g_1\, , \, g_2^{-1}k_2g_2} \delta_{g_2^{-1}k_2g_2\, , \,g_3^{-1}k_3g_3}
\Big({}^{\wp_1 \, \wp_2 \, \wp_3}_{O_1 O_2 O_3}\Big)_{\rm I}
\end{align}
where we used the defining property of the representations of the quantum triple. By comparing the first and the last line of \eqref{inv_Ex2}, we conclude that the $3 \wp M$-symbols implicitly encode the zero-flux condition on the surface of $\mathbb{Y} \times \mathbb{S}_1$, the gauge invariance at the single bulk vertex, as well as the identification of the $\{g^{-1}kg\}$-holonomies along the $\mathbb{S}_1$ direction.

Using the states \eqref{fusThrice3D}, we can construct the fusion basis for excited states defined on manifolds of the form $\Sigma \times \mathbb{S}_1$. To do so, we rely upon the fact that the manifold $\Sigma \times \mathbb{S}_1$ can be obtained as a sewing of several copies of $\mathbb{Y} \times \mathbb{S}_1$. The strategy is to perform such a decomposition of the manifold, associate a state $|\{\wp_i,M_i\}_{i=1}^3 \ra_{\mathbb{Y} \times \mathbb{S}_1}$ to each copy of $\mathbb{Y} \times \mathbb{S}_1$, and contract them to each other following the decomposition pattern. Equivalently, we can associate an intertwining map $\mathcal{I}^{\wp_1\wp_2\wp_3}$ to each copy of $\mathbb{Y} \times \mathbb{S}_1$ and contract them via three-cylinder fusion basis states $|\wp,MN \ra_{\mathbb{I}_3}$. For a given manifold $\Sigma \times \mathbb{S}_1$, and for a given decomposition $\{\mathbb{Y}\times \mathbb{S}_1\}$, a formal expression for the corresponding fusion basis states therefore reads
\be
\boxed{
	\big| \{\wp\}\big\ra_{\Sigma \times \mathbb{S}_1} = 
	{\rm tr}_{\{V^{\wp}\}}\Big[\bigotimes_{\mathbb{I}_3}|\wp \ra_{\mathbb{I}_3}
	\otimes \bigotimes^{ }_{\mathbb{Y} \times \mathbb{S}_1}\mathcal{I}^{\{\wp\}} \Big] 
\; .}
\ee
The fusion basis is {\it orthogonal} and {\it complete}. This follows directly from the orthogonality \eqref{triple_ortho} and completeness \eqref{triple_complete} of the representation matrices as well as the orthogonality \eqref{ortho_CG3D} and completeness \eqref{complete_CG3D} of the $3 \wp M$-symbols coefficients. 
\newpage

\section{Discussion}
Although the fusion basis for the three-torus, as presented in this paper, appeared before, see {\it e.g.} \cite{Wan:2014woa, Wang:2014oya}, the corresponding algebraic structure was yet to be explored. By following the strategy employed in (2+1)d to reveal the Drinfel'd double, we discovered this algebraic structure, namely the quantum triple $\mT(G)$. In addition, we showed explicitly how the ground states on the three-torus are in one-to-one correspondence with the quasi-excitations defined on the manifold obtained by cutting the three-torus along one direction. 

Furthermore, we presented a method to define the fusion basis for general excited states. In this construction, excitations are restricted to happen at boundary two-tori such that we are dealing with manifolds of the form $\Sigma \times \mathbb{S}_1$. The definition of the fusion basis relies upon the fact that such manifolds can be obtained as the sewing of several copies of $\mathbb{Y} \times \mathbb{S}_1$, namely the direct product between the thrice-punctured two-sphere and the circle. Such decomposition then dictates a simple way of constructing the basis: To each copy of $\mathbb{Y} \times \mathbb{S}_1$ we assign an intertwining map, which we contract to each other via three-cylinder basis states. The resulting states are labeled by two sets of conjugacy classes $\{C\}$ and $\{D\}$, representing fluxes, and a set of irreducible representations $\{R\}$, representing charges. The definition of the fusion basis is tied to the choice of comultiplication maps when constructing the quantum triple. Here we made a natural choice such that the fusion category of representations of the quantum triple $\mT(G)$ reduces to fusion categories of representations of Drinfel'd doubles $\mD(Z_C)$. As such, it turns out that our construction provides an algebraic translation of the dimensional reduction strategy. 

We conjecture that defining a similar trialgebra associated with different choices of coalgebras would lead to different fusion bases. These alternative bases would correspond to different fusion patterns of the manifold boundaries hence yielding more general topologies. Alternatively, we could reproduce the construction presented in this paper by replacing the three-cylinder with another manifold of the form $\Sigma \times I$. This would lead to another version of the 3d tube algebra yielding yet another algebraic structure.

The (2+1)d fusion basis associated to a given punctured surface, diagonalizes a set of closed ribbon operators \cite{Kitaev:1997wr, Bombin:2007qv, DDR1}. These operators, which measure the excitation content of a given region, are constructed as a composition of Wilson loop operators and parallel-transported translation operators. We expect that the fusion basis for (3+1)d topological phases we propose in this paper also diagonalizes a set of analogous operators. These operators should be an extension of the ribbon operators in the same way as the quantum triple is an extension of the Drinfel'd double.

We focused our study on the case where the ground state is described by a $BF$ theory. The natural next step of this work would be to generalize to Dijkgraaf-Witten theory which can be thought as a twisted $BF$ theory such that the twist deforms the Gau{\ss} constraint. The twisted case differs in particular from the non-twisted case in the definition of the local equivalence relations. In (2+1)d, when picking the graph to be the one-skeleton of a triangulation, the local equivalence relations can be defined in terms of Pachner moves. In particular, the 2-2 Pachner move is performed by a map which evaluates to a group 3-cocycle $\alpha$ in $H^3(G,{\rm U}(1))$. The tube algebra then requires three Pachner moves so that the multiplication is deformed by a phase
\begin{equation*}
		\theta_{h}(g_1,g_2) = \frac{\alpha(h,g_1,g_2)\alpha(g_1,g_2, (g_1g_2)^{-1}hg_1g_2)}{\alpha(g_1,g_1^{-1}hg_1,g_2)} \; .
\end{equation*}  
It turns out that this phase is the slant product $i_h\alpha$ which pairs the group element $h$ with the 3-cocycle $\alpha$. Algebraically, this corresponds to turning the Drinfel'd double into a quasi-Hopf algebra whose twist reads
\begin{equation*}
	\phi = \sum_{h_1,h_2,h_3}\alpha(h_1,h_2,h_3)^{-1}(\mathbbm{1}_G \smo \delta_{h_1})
	\otimes (\mathbbm{1}_G \smo \delta_{h_2})\otimes (\mathbbm{1}_G \smo \delta_{h_3}) \; .
\end{equation*}
The same strategy generalizes to the three-cylinder algebra. Roughly speaking, the key is to realize that when working with a manifold of the form $\Sigma \times \mathbb{S}_1$, the (3+1)d construction mimics the (2+1)d one where the group 3-cocycle $\alpha$ is now replaced by the slant product $i_k \omega$ where $\omega$ is an element of $H^4(G,{\rm U}(1))$ and $k$ is the holonomy along the compactified direction. 
 In general, the result of the slant product $i_k \omega$ satisfies a so-called \emph{twisted 3-cocycle condition}. However, thanks to the commutativity between $h$ and $k$ holonomies, it actually satisfies the usual group 3-cocycle condition. This means that an associator can be defined, as for the Drinfel'd double, and can be used in order to `twist' the quantum triple. It implies that the multiplication and the comultiplications are deformed by twisted 2-cocyles obtained as slant products of $i_k\omega$. For instance, the twisted 2-cocyle deforming the multiplication rule is given by
\begin{equation*}
	\theta_{h,k}(g_1,g_2) = \frac{i_k\omega(h,g_1,g_2)i_k\omega(g_1,g_2, (g_1g_2)^{-1}hg_1g_2)}{i_k\omega(g_1,g_1^{-1}hg_1,g_2)}
\end{equation*}
so that the multiplication rule now reads
\begin{equation*}
	(g_1 \smo \delta_{h_1} \smoc \delta_{k_1})\star (g_2 \smo \delta_{h_2} \smoc \delta_{k_2}) := \delta_{k_2,g_1^{-1}k_1g_1}\delta_{h_2,g_1^{-1}h_1g_1} \theta_{h_1,k_1}(g_1,g_2)(g_1g_2 \smo \delta_{h_1} \smoc \delta_{k_1})
\end{equation*}
This generalization should be particularly interesting since it is believed that (3+1)d bosonic topological orders with bosonic point-like excitations are classified by a pair $(G, \omega)$, with $G$ a finite group and $\omega$ a group 4-cocycle \cite{2017arXiv170404221L}. We postpone a deeper study of this model to another paper.

\acknowledgments
It is a pleasure to thank Bianca Dittrich and Aldo Riello for a careful reading of this manuscript and for many insightful comments. We also thank Christophe Goeller and Theo Johnson-Freyd for usedul discussions.
CD is supported by an NSERC
grant awarded to Bianca Dittrich. This research was supported in part by Perimeter Institute for Theoretical Physics. Research
at Perimeter Institute is supported by the Government of Canada through the Department of Innovation, Science and
Economic Development Canada and by the Province of Ontario through the Ministry of Research, Innovation and
Science.

\newpage
\appendix

\section{Properties of the irreducible representations of the quantum triple \label{app_irrepsandsoon}}
\vspace{-0.2em}
\subsection{Defining property of the representations \label{app_defprop}}
The irreducible representations of the quantum triple are homomorphisms and as such they preserve the algebraic structure. The following shows how the irreducible representations are compatible with the multiplication rule $\star$:

\begin{align} \nn
	&\sum_{b,\beta,n}D^{\wp}_{a\alpha m ,b \beta n}(g_1 \smo \delta_{h_1} \smo \delta_{k_1})D^{\wp}_{b \beta n ,c \gamma o}(g_2 \smo \delta_{h_2} \smo \delta_{k_2}) \\[-0.1em] \nn
	& \q = \sum_{b,\beta,n}
	\delta(k_1,c_a)\delta(c_a,g_1c_bg_1^{-1})\delta(h_1k_1,k_1h_1)\delta(d_\alpha, p_a^{-1}g_1p_bd_\beta p_b^{-1}g_1^{-1}p_a)\delta(p_a^{-1}h_1p_a,d_\alpha) \\[-0.6em] \nn
	& \q \q \q \times 
	\delta(k_2,c_b)\delta(c_b,g_2c_cg_2^{-1})\delta(h_2k_2,k_2h_2)\delta(d_\beta, p_b^{-1}g_2p_cd_\gamma p_c^{-1}g_2^{-1}p_b)\delta(p_b^{-1}h_2p_b,d_\beta) \\[0.4em] \nn
	& \q \q \q \times D^R_{mn}(q_\alpha^{-1}p_a^{-1}g_1p_bq_\beta)
	D^R_{no}(q_\beta^{-1}p_b^{-1}g_2p_cq_\gamma) \\[0.3em] \nn
	& \q = \delta(k_1,c_a)\delta(c_a,g_1g_2c_cg_2^{-1}g_1^{-1})\delta(h_1k_1,k_1h_1)
	\delta(d_\alpha,p_a^{-1}g_1g_2p_cd_\gamma p_c^{-1}g_2^{-1}g_1^{-1}p_a) 
	\delta(d_\alpha,p_a^{-1}h_1p_a)\\[0.2em] \nn
	& \q \q \q \times 
	\underbrace{\delta(k_2,g_2c_cg_2^{-1})}_{\delta(k_2,g_1^{-1}k_1g_1)}
	\underbrace{\delta(h_2,g_2p_cd_\gamma p_c^{-1}g_2^{-1})}_{\delta(h_2,g_1^{-1}h_1g_1)}
	D^R_{mo}(q_\alpha^{-1}p_a^{-1}g_1g_2p_cq_\gamma) \\[0.3em]
	& \q = D^{\wp}_{a\alpha m , c \gamma o }((g_1 \smo \delta_{h_1} \smo \delta_{k_1}) \star (g_2 \smo \delta_{h_2} \smo \delta_{k_2}))
\end{align}
where we used the fact that the irreducible representations of the stabilizer $Z_D$ preserve the group multiplication rule.

\subsection{Orthogonality of the irreducible representations}\label{app_ortho}

The space of functions on $\mT(G)$ is equipped with an inner product defined by
\be
	\langle \psi, \phi \rangle = \frac{1}{|G|}\sum_{g,h,k \in G}\overline{\psi(g \smo \delta_h \smo \delta_k)}\phi(g \smo \delta_h \smo \delta_k) \; .
\ee
The matrix elements of the irreducible representations of the quantum triple form an orthogonal set with respect to this inner product, i.e.
\begin{align} \nn
	&\frac{1}{|G|}\sum_{g,h,k \in G}\overline{D^{\wp}_{a\alpha m , b \beta n }(g \smo \delta_h \smo \delta_k)}D^{\widetilde{\wp}}_{\wa \walpha \wm , \wb \wbeta \wn }(g \smo \delta_h \smo \delta_k) \\ \nn
	& \q  = \frac{1}{|G|}\sum_{g,h,k \in G}
	\delta(k,c_a)\delta(c_a,gc_bg^{-1})\delta(hk,kh)\delta(p_a^{-1}hp_a,d_\alpha)
	\delta(d_\alpha,p_a^{-1}gp_bd_\beta p_b^{-1}g^{-1}p_a) \\[-0.3em] \nn
	& \hspace{5.8em}\times 
	\delta(k,\wc_{\wa})\delta(\wc_{\wa},g\wc_{\wb}g^{-1})\delta(\wip_{\wa}^{-1}h\wip_{\wa},\wid_{\walpha})
	\delta(\wid_{\walpha},\wip_{\wa}^{-1}g\wip_{\wb}\wid_{\wbeta} \wip_{\wb}^{-1}g^{-1}\wip_{\wa}) \\[0.4em] \nn
	& \hspace{5.8em} \times
	\overline{D^R_{mn}(q_{\alpha}^{-1}p_a^{-1}gp_bq_\beta)}
	D^{\widetilde{R}}_{\wm \wn}(\wq_{\walpha}^{-1}\wip_{\wa}^{-1}g\wip_{\wb}\wq_{\wbeta})\\[0.2em] \nn
	& \q = \frac{1}{|C||D||Z_D|} \sum_{g\in G}
	\delta_{C,\wC}\delta_{a,\wa}\delta_{b,\wb}\delta_{D,\widetilde{D}}\delta_{\alpha,\walpha}
	\delta_{\beta,\wbeta}
	\delta(c_a,gc_bg^{-1})\delta(d_\alpha,p_a^{-1}gp_bd_\beta p_b^{-1}g^{-1}p_a) \\ \nn
	&\hspace{5.8em} \times \overline{D^R_{mn}(q_{\alpha}^{-1}p_a^{-1}gp_bq_\beta)}
	D^{\widetilde{R}}_{\wm \wn}(q_{\alpha}^{-1}p_a^{-1}gp_bq_\beta) \\ 
	& \q = \frac{\delta_{C,\wC}\delta_{D,\widetilde{D}}\delta_{R,\wR}}{|C||D|d_R}
	\delta_{a,\wa}\delta_{b,\wb}\delta_{\alpha,\walpha}\delta_{\beta,\wbeta}
	\delta_{m,\wm}\delta_{n,\wn} = \frac{\delta_{\wp,\widetilde{\wp}}}{d_\wp}\delta_{M,\widetilde{M}}\delta_{N,\widetilde{N}}
\end{align}
where we used between the last two lines the orthogonality of the irreducible representations of the stabilizer $Z_{C}$ and the fact that the cardinality of the group $G$ decomposes as $|G|=|C|\cdot |D| \cdot |Z_D|$.

\subsection{Completeness of the set of irreducible representations}
The irreducble representations form a complete set of representations, since they resolve the identity on $\mT(G)$:

\begin{align} \nn
	&\sum_{\wp}\sum_{a,\alpha,m \atop b,\beta,n}d_{\wp}\overline{D^\wp_{a \alpha m, b \beta n}(g_1 \smo \delta_{h_1} \smo \delta_{k_1})}
	D^{\wp}_{a \alpha m , b \beta n }(g_2 \smo \delta_{h_2} \smo \delta_{k_2}) \\ \nn
	& \q = 
	\sum_{\wp}\sum_{a,\alpha,m \atop b,\beta,n}
	\delta(k_1,c_a)\delta(c_a,g_1c_bg_1^{-1})\delta(h_1k_1,k_1h_1)\delta(p_a^{-1}h_1p_a,d_\alpha)
	\delta(d_\alpha,p_a^{-1}g_1p_bd_\beta p_b^{-1}g_1^{-1}p_a) \\[-0.5em] \nn
	& \hspace{5em} \times
	\delta(k_2,c_a)\delta(c_a,g_2c_bg_2^{-1})\delta(h_2k_2,k_2h_2)\delta(p_a^{-1}h_2p_a,d_\alpha)
	\delta(d_\alpha,p_a^{-1}g_2p_bd_\beta p_b^{-1}g_2^{-1}p_a) \\[0.5em] \nn
	& \hspace{5em} \times |C||D|d_R
	\overline{D^R_{mn}(q_{\alpha}^{-1}p_a^{-1}g_1p_bq_\beta)}
	D^R_{mn}(q_{\alpha}^{-1}p_a^{-1}g_2p_bq_\beta) \\[0.4em] \nn
	& \q = |G| \sum_{C,D}\sum_{a, \alpha \atop b, \beta}
	\delta(k_1,c_a)\delta(c_a,g_1c_bg_1^{-1})\delta(h_1k_1,k_1h_1)\delta(p_a^{-1}h_1p_a,d_\alpha)\delta(d_\alpha,p_a^{-1}g_1p_bd_\beta p_b^{-1}g_1^{-1}p_a) \\[-0.5em] \nn
	& \hspace{5.8em} \times \delta(g_1,g_2)\delta(k_2,c_a)\delta(h_2k_2,k_2h_2)\delta(p_a^{-1}h_2p_a,d_\alpha) \\[0.2em]
	& \q = |G|\delta_{g_1,g_2}\delta_{h_1,h_2}\delta_{k_1,k_2}\delta_{h_1k_1,k_1h_1}
\end{align}
where we used the completeness of the set of Wigner matrices. 

\subsection{Invariance property of the $3 \wp M$-symbols \label{app_invCG}}

Let us focus on the case of the $3 \wp M$-symbols defined with respect to the comultiplciation $\Delta_{\rm I}$. The defining equation for the Clebsch-Gordan coefficients ${}_{\rm I}\mC$ of the quantum triple reads

\begin{align}
	D^{\wp_1}_{M_1N_1} \otimes_{\rm I} D^{\wp_2}_{M_2N_2}(\Delta_{\rm I}(g \smo \delta_h \smo \delta_k))
	\,=\, 
	\sum_{\wp_3}\sum_{M_3,N_3}
	{}_{\rm I}\mathcal{C}^{\wp_1\wp_2\wp_3}_{M_1M_2M_3}\, D^{\wp_3}_{M_3N_3}(g \smo \delta_h \smo \delta_k)\,\overline{{}_{\rm I}\mathcal{C}^{\wp_1\wp_2\wp_3}_{N_1N_2N_3}}
\end{align}
which, using the unitarity of the map ${}_{\rm I}\mC^{\wp_1 \wp_2}$, can be rewritten 
\begin{align}
	\label{Inv2}
	\sum_{N_1,N_2}D^{\wp_1}_{M_1N_1} \otimes_{\rm I} D^{\wp_2}_{M_2N_2}(\Delta_{\rm I}(g \smo \delta_h \smo \delta_k))  \,{}_{\rm I}\mC^{\wp_1 \wp_2 \wp_3}_{N_1 N_2 N_3}
	&=  \sum_{M_3}
	{}_{\rm I}\mC^{\wp_1 \wp_2 \wp_3}_{M_1 M_2 M_3}\,
	D^{\wp_3}_{M_3N_3}(g \smo \delta_h \smo \delta_k) \; .
\end{align}
We can now make use of the equation
\begin{align} \nn
	&\sum_{N_3}\sum_{h,k \in G} D^{\wp_3}_{M_3N_3}(g \smo \delta_h \smo \delta_k)  D^{\wp_3}_{N_3 O_3}(g^{-1} \smo \delta_{g^{-1}hg} \smo \delta_{g^{-1}kg}) \\
	& \q =  
	\sum_{h,k \in G}   D^{\wp_3}_{M_3O_3}(\mathbbm{1}_G \smo \delta_h \smo \delta_k) = D^{\wp_3}_{M_3O_3}( \mathbbm{1}_{\mT(G)}) = \delta_{M_3 O_3}
\end{align}
by multiplying (\ref{Inv2}) from the right with $D^{\wp_3}_{N_3O_3}(g^{-1} \smo \delta_{g^{-1}hg} \smo \delta_{g^{-1}kg})$ and summing over $h,k$. Resolving the comultiplication and remembering the definition \eqref{def_dualrep} of the dual representation $\wp^{\ast}$, we finally obtain the invariance property of the Clebsch-Gordan coefficients:
\begin{align}
	&{}_{\rm I}\mC^{\wp_1\wp_2\wp_3}_{M_1M_2M_3} = 
	\sum_{h_1,h_2 \atop k}
	D^{\wp_1}_{M_1N_1}(g \smo \delta_{h_1} \smo \delta_{k})
	D^{\wp_2}_{M_2N_2}(g \smo \delta_{h_2} \smo \delta_{k})
	D^{\wp_3^{\ast}}_{M_3N_3}(g \smo \delta_{h_2^{-1}h_1^{-1}} \smo \delta_{k})\, {}_{\rm I}\mC^{\wp_1\wp_2\wp_3}_{N_1N_2N_3} 
\end{align}
from which, it is straightforward to deduce the invariance property of the $3 \wp M$-symbols
\begin{align}
	\Big({}^{\, \wp_1 \;\, \wp_2 \;\, \wp_3}_{M_1M_2M_3}\Big)_{\rm I}  &= 
	\sum_{h_1,h_2 \atop k}
	D^{\wp_1}_{M_1N_1}(g \smo \delta_{h_1} \smo \delta_{k})
	D^{\wp_2}_{M_2N_2}(g \smo \delta_{h_2} \smo \delta_{k})
	D^{\wp_3}_{M_3N_3}(g \smo \delta_{h_2^{-1}h_1^{-1}} \smo \delta_{k}) 	\Big({}^{\, \wp_1 \; \wp_2 \; \wp_3}_{N_1N_2N_3}\Big)_{\rm I} .
\end{align}

\bibliographystyle{JHEP}
\bibliography{ExcBasis}

\end{document}